\documentclass[reprint,showpacs,preprintnumbers,amsmath,amssymb, citeautoscript,prl,superscriptaddress]{revtex4-2}
\usepackage{graphicx,wasysym}
\usepackage{dcolumn}
\usepackage{bm,mathtools}

\setcitestyle{super}
\newcommand*{\citen}[1]{%
  \begingroup
    \romannumeral-`\x 
    \setcitestyle{numbers}%
    \cite{#1}%
  \endgroup   
}

\usepackage[hypertexnames=false]{hyperref}

\usepackage{tikz}
\usetikzlibrary{positioning}
\usetikzlibrary{arrows.meta}
\usepackage{xcolor}
\usepackage{algorithm}
\usepackage{algpseudocode}

\makeatletter
\def\maketitle{
\@author@finish
\title@column\titleblock@produce
\suppressfloats[t]}
\makeatother

\begin{document}
\title{Exact computation of Transfer Entropy with Path Weight Sampling}

\preprint{APS/123-QED}
\author{Avishek Das}
\email{a.das@amolf.nl}
\affiliation{AMOLF, Science Park 104, 1098 XG, Amsterdam, The Netherlands\looseness=-1}
\author{Pieter Rein ten Wolde}
\email{p.t.wolde@amolf.nl}
\affiliation{AMOLF, Science Park 104, 1098 XG, Amsterdam, The Netherlands\looseness=-1}

\date{\today}
\begin{abstract}
The ability to quantify the directional flow of information is vital to understanding natural systems and designing engineered information-processing systems. A widely used measure to quantify this information flow is the transfer entropy. However, until now, this quantity could only be obtained in dynamical models using approximations that are typically uncontrolled. 
Here we introduce a computational algorithm called Transfer Entropy-Path Weight Sampling (TE-PWS), which makes it possible, for the first time, to quantify the transfer entropy and its variants exactly for any stochastic model, including those with multiple hidden variables, nonlinearity, transient conditions, and feedback. By leveraging techniques from polymer and path sampling, TE-PWS efficiently computes the transfer entropy as a Monte-Carlo average over signal trajectory space. 
We use our exact technique to demonstrate that commonly used approximate methods to compute transfer entropies incur large systematic errors and high computational costs.
As an application, we use TE-PWS in linear and nonlinear systems to reveal how transfer entropy can overcome naive applications of the data processing inequality in the presence of feedback.
\end{abstract}
\maketitle

\thispagestyle{empty}
\twocolumngrid

Information transfer between noisy signals underlies the functionality of diverse natural and man-made networks such as those in biochemical signaling, neuroscience, ecology, wireless communication and finance. Information theory has so far provided a useful framework for quantifying information transmission. In the presence of feedback loops in the network, information travels in both directions between the input and the output. An information-theoretic measure that can quantify the information transfer separately in either direction is the transfer entropy\cite{schreiber2000measuring,spinney2017transfer}. Transfer entropy and its variants, such as directed information, conditional transfer entropy or filtered transfer entropy, have been widely used to gain knowledge about the connectivity of a network\cite{vicente2011transfer}, infer causal relations from experiments\cite{runge2018causal,hlavavckova2007causality}, establish fundamental bounds on network performance\cite{mattingly2021escherichia}, and estimate the minimal physical work required for a computation\cite{prokopenko2014transfer,horowitz2014second}. Hence, for a wide range of problems, it is vital to be able to accurately quantify transfer entropies.

However, there are currently no exact methods to compute the transfer entropy in a general many-variable dynamical model. The transfer entropy depends on probability distributions of signal trajectories. Estimating these distributions by binning the experimental or simulated trajectories in histograms is not feasible, as the dimensionality of the trajectory space scales exponentially with signal duration\cite{strong1998entropy,runge2012escaping}. As a result, different approximations are currently being used: either the full history dependence of the transfer entropy is truncated\cite{lahiri2017information}, or an arbitrary distance metric in trajectory space is chosen for clustering the trajectories\cite{shorten2021estimating}, or a linear or low-order moment-closure approximation is employed\cite{tostevin2009mutual,barnett2014mvgc,moor2023dynamic}. These approximations can result in uncontrolled errors in complex, nonlinear, many-variable systems\cite{potter2017dynamic,cepeda2019estimating}. In the absence of an exact method, the magnitude of these errors remains unclear.

Here we fill this gap by introducing TE-PWS, a numerical algorithm to estimate transfer entropies exactly for any stochastic model, including diffusive and jump processes. The estimate is exact, \textit{i.e.}, it is an unbiased statistical estimate of the transfer entropy. TE-PWS can therefore provide \textit{ground truth} results for any given model. TE-PWS builds on the recently developed PWS algorithm for computing the mutual information between trajectories\cite{reinhardt2023path}. TE-PWS exploits the idea that path likelihoods can be obtained analytically from the Langevin or master equation, from which the transfer entropy is then computed via Monte-Carlo averaging in trajectory space. Additionally, long trajectories are sampled with an importance sampling scheme\cite{frenkel2019understanding}, solving the problem of exponential scaling of the computational cost with trajectory duration. 
We first show that TE-PWS reproduces analytical results when available.
We then apply TE-PWS to compute the transfer entropy in a three-variable motif in the presence of feedback, for both linear and nonlinear systems, yielding novel insights on how information feedback can amplify information transfer. Specifically, the transfer entropy from an input to an output node can overcome a naive application of the data processing inequality even when the mutual information obeys one.
Finally, we use TE-PWS to show that the most widely used approximate methods for the computation of transfer entropy--- the Gaussian framework, KSG and single-step truncation--- all exhibit significant systematic errors. The comparison also reveals that TE-PWS is computationally efficient.

{\bf Transfer entropy.} 
The TE-PWS algorithm has the same key steps for all stochastic models. We therefore describe TE-PWS for diffusive processes in the main text and for jump processes in the End Matter (EM). Consider a $d$-dimensional diffusive process $\mathbf{X}(t)$ modelled as a function of time $t$ by a Langevin equation
\begin{align}\label{eq:langevin}
\dot{\mathbf{X}}(t)=\mathbf{F}(t)+\bm{\xi}(t),
\end{align}
with $\mathbf{F}(t)$ a general drift, and $\bm{\xi}(t)$ a $d$-dimensional Gaussian white noise with a diffusion constant matrix $\mathbf{D}=[D_{ij}]$ such that $\langle\xi_{i}(t)\xi_{j}(t^{'})\rangle=2D_{ij}\delta(t-t^{'})$. The drift may depend on the entire past history as well as on time. The transfer entropy from $X_{i}$ to $X_{j}$ over $N$ timesteps of durations $\delta t$ each is defined as\cite{schreiber2000measuring}
\begin{align}
\mathcal{T}_{X_{i}\to X_{j}}&=\sum_{k=0}^{N-1}I\left( X_{j}(k+1);X_{i,[0,k]}|X_{j,[0,k]}\right) \label{eq:transfer}\\
&\equiv\sum_{k=0}^{N-1}\mathcal{T}_{X_{i}\to X_{j}}^{[k]}\label{eq:kdependentsum}
\end{align}
where the index $k$ goes over individual timesteps, $X_{j}(k)$ denotes $X_{j}$ after $k$ timesteps, $X_{j,[0,k]}$ denotes the trajectory over the first $k$ timesteps, $\mathcal{T}_{X_{i}\to X_{j}}^{[k]}$ denotes the $k$-th term in the sum, and $I(A;B)$ denotes the mutual information between two random variables $A$ and $B$. $\mathcal{T}_{X_{i}\to X_{j}}$ measures the information transferred from the past trajectory of $X_{i}$ to the new updates of $X_{j}$ at every timestep, given the past trajectory of $X_{j}$ is already known. In case the dynamics is in steady-state, we will also talk about the transfer entropy rate, $\dot{\mathcal{T}}_{X_{i}\to X_{j}}=\lim_{N\to\infty}\mathcal{T}_{X_{i}\to X_{j}}/(N\delta t)$.

\begin{figure}[t]
\centering
\includegraphics[width=8.6cm]{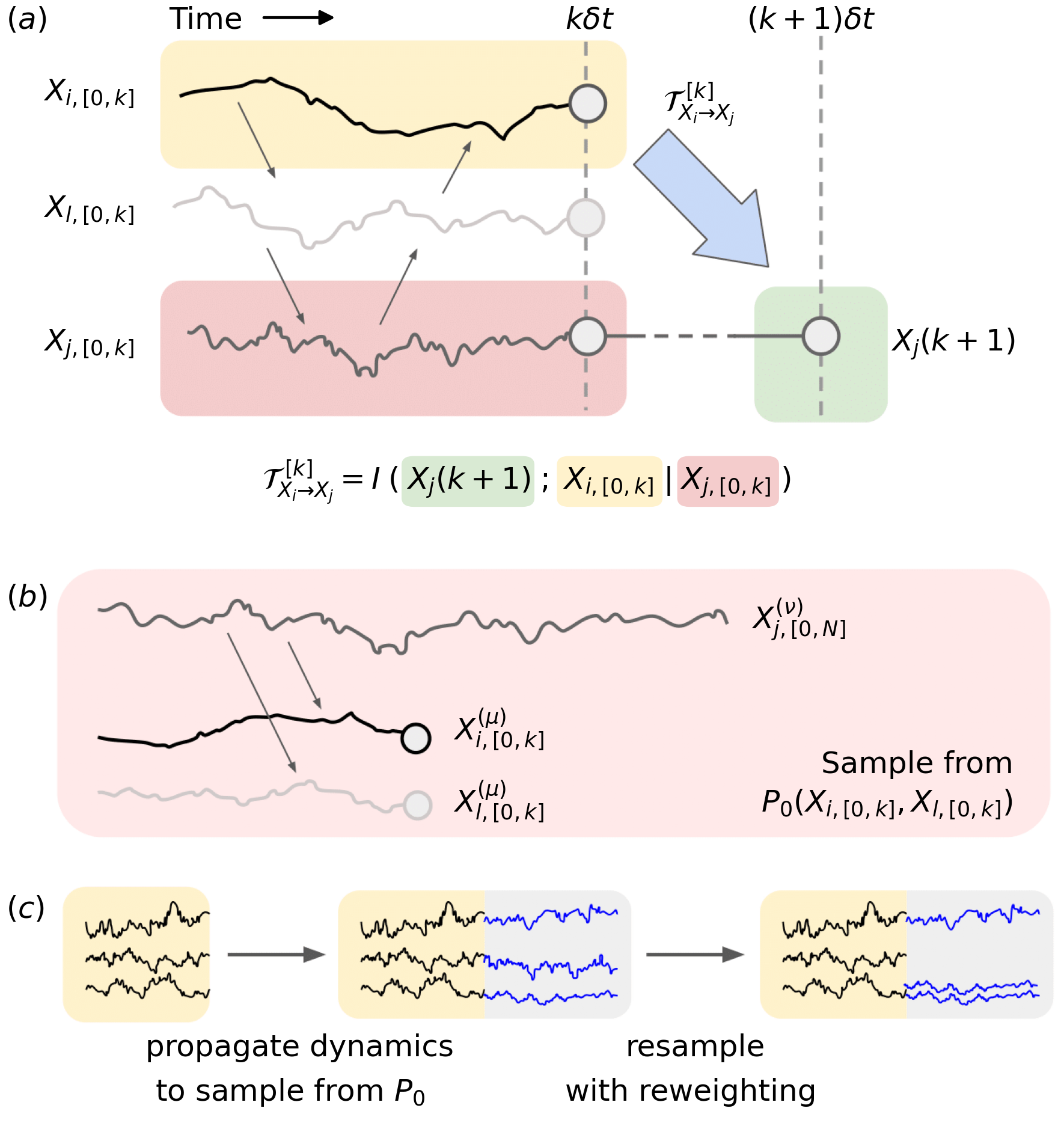}
\caption{(a) Schematic representation of the increase in transfer entropy from variable $X_{i}$ to $X_{j}$ at the $(k+1)$-th timestep. Other variables $X_{l}$ may mediate information transfer even in the absence of a direct coupling from $X_{i}$ to $X_{j}$. (b) Propagation of reference dynamics for $X_{i}$ and $X_{l}$ such that it is commensurate with the given frozen $X_{j}^{(\nu)}$ trajectory.
(c) In the RR scheme, trajectories sampled from a reference distribution $P_{0}(X_{i,[0,k]},X_{l,[0,k]})$ are resampled periodically to turn them into the desired conditional distribution $P(X_{i,[0,k]},X_{l,[0,k]}|X_{j,[0,k]}^{(\nu)})$.}
\label{fig:schematic}
\end{figure}

We can rewrite the transfer entropy equivalently as
\begin{align}
\mathcal{T}_{X_{i}\to X_{j}}&=\sum_{k=0}^{N-1}\mathcal{T}_{X_{i}\to X_{j}}^{[k]}=\sum_{k} H\left( X_{j}(k+1)| X_{j,[0,k]}\right) \nonumber\\
&\qquad \qquad- H\left( X_{j}(k+1)|X_{i,[0,k]},X_{j,[0,k]}\right) \label{eq:transfer1}\\
=&\sum_{k}\Bigg\langle \ln\frac{P\left( X_{j}(k+1)|X_{i,[0,k]},X_{j,[0,k]}\right) }{P\left( X_{j}(k+1)|X_{j,[0,k]}\right) } \Bigg\rangle \label{eq:transferentropy_probab}
\end{align}
where $H(A)$ denotes the Shannon entropy associated with the probability distribution $P(A)$ of $A$, and the angular brackets denote an average over the joint probability $P\left( X_{i,[0,N]},X_{j,[0,N]}\right) $.
Eq. \ref{eq:transfer1} shows that transfer entropy quantifies the additional information in $X_{j}(k+1)$ that arrives from $X_{i,[0,k]}$ beyond that which is already present in the past trajectory $X_{j,[0,k]}$. This occurs either through direct causal action, or through a third variable $X_{l}$, schematically demonstrated in Fig. \ref{fig:schematic}a. If $X_{i}$ does not affect the dynamics of $X_{j}$, this additional information would be zero. In general, every $\mathcal{T}_{X_{i}\to X_{j}}^{[k]}$ is a mutual information and hence nonnegative.

For calculating the transfer entropy using Eq. \ref{eq:transferentropy_probab} we develop TE-PWS. Following PWS\cite{reinhardt2023path}, the central idea is that trajectory likelihoods in the full $d$-dimensional space are analytically available on-the-fly, and that trajectory averages can be computed in a Monte-Carlo fashion. First, the average in Eq. \ref{eq:transferentropy_probab} is computed as,
\begin{align}\label{eq:temcestimate1}
\mathcal{T}_{X_{i}\to X_{j}}&=\frac{1}{M_{1}}\sum_{\nu}\sum_{k}\ln\frac{P\left( X^{(\nu)}_{j}(k+1)\big| X^{(\nu)}_{i,[0,k]},X^{(\nu)}_{j,[0,k]}\right) }{P\left( X^{(\nu)}_{j}(k+1)\big| X^{(\nu)}_{j,[0,k]}\right) }
\end{align}
where the index $\nu$ sums over $M_{1}$ pairs of trajectories of $X_{i}$ and $X_{j}$ sampled from the joint probability distribution $P\left( X_{i,[0,N]},X_{j,[0,N]}\right)$. The superscript $(\nu)$ in $X_{i}^{(\nu)}$ refers to the $\nu$-th statistical realization of $X_{i}$. For each pair of trajectories, the probabilties in the numerator and the denominator of Eq. \ref{eq:temcestimate1} are not analytically available, but what is indeed available is the full joint probability $P(\mathbf{X}_{[0,N]}^{(\nu)})$ as the exponential of the Onsager-Machlup action\cite{rogers2000diffusions,onsager1953fluctuations}. We thus need to marginalize over all degrees of freedom other than $X_{i}$ and $X_{j}$, denoted henceforth collectively as $X_{l}$. We illustrate this procedure first for the denominator in Eq. \ref{eq:temcestimate1}. It is obtained as $P(X^{(\nu)}_{j,[0,k+1]})/P(X^{(\nu)}_{j,[0,k]})$, where $P(X^{(\nu)}_{j,[0,k]})$ is obtained via marginalization,
\begin{align}\label{eq:denominator}
&P\left( X_{j,[0,k]}^{(\nu)}\right) \nonumber\\
&=\int\int D[X_{i,[0,k]}]D[X_{l,[0,k]}]P\left( X_{i,[0,k]},X_{j,[0,k]}^{(\nu)},X_{l,[0,k]}\right) 
\end{align}
For performing this average in a Monte-Carlo fashion\cite{frenkel2019understanding}, we sample from a reference distribution, $P_{0}(X_{i,[0,k]},X_{l,[0,k]})$, and correct the resultant bias by dividing by $P_{0}$,
\begin{align}
P\left( X_{j,[0,k]}^{(\nu)}\right) &=\frac{1}{M_{2}}\sum_{\mu} \frac{P\left( X_{i,[0,k]}^{(\mu)},X_{j,[0,k]}^{(\nu)},X_{l,[0,k]}^{(\mu)}\right) }{P_{0}\left( X_{i,[0,k]}^{(\mu)},X_{l,[0,k]}^{(\mu)}\right) }\label{eq:temcestimate2}
\end{align}
where the index $\mu$ sums over $M_{2}$ trajectories sampled from $P_{0}$.
What is the best choice for $P_{0}$? The ideal choice would be the conditional distribution $P(X_{i,[0,k]},X_{l,[0,k]}|X_{j,[0,k]}^{(\nu)})$, as it makes the summand in Eq. \ref{eq:temcestimate2} equal for all $\mu$, such that the variance of the estimate of $P( X_{j,[0,k]}^{(\nu)})$ is zero. However, this conditional distribution is not known \textit{a priori}. We therefore generate $X_{i}$ and $X_{l}$ trajectories in the frozen field of $X_{j,[0,N]}^{(\nu)}$ (Fig. \ref{fig:schematic}b) resulting in a distribution $P_{0}$ that is known analytically and is expected to be close to this conditional distribution. To exactly compensate for the remaining deviations of $P_{0}$ from the desired conditional distribution, we employ, in the spirit of Rosenbluth-Rosenbluth(RR)-PWS\cite{reinhardt2023path}, a reweighing of the $X_{i}^{(\mu)}$ and $X_{l}^{(\mu)}$ trajectories on-the-fly with weights proportional to the ratio of the two distributions. We resample the trajectories with these weights after every $\delta t$ time, meaning that we sample $M_{2}$ trajectories with replacement from the weighted trajectory ensemble (Fig.~\ref{fig:schematic}c). This procedure exactly generates $X_{i}^{(\mu)}$ and $X_{l}^{(\mu)}$ trajectories according to the desired conditional distribution $P(X_{i,[0,k]},X_{l,[0,k]}|X_{j,[0,k]}^{(\nu)})$ (see EM and Supplemental Material (SM) section SM-C).

Returning now to the numerator in Eq. \ref{eq:temcestimate1}, we note that it is an average over another conditional distribution,
\begin{align}
&\hspace{0.2in}P\left( X_{j}^{(\nu)}(k+1)\big|X_{i,[0,k]}^{(\nu)},X_{j,[0,k]}^{(\nu)}\right) \nonumber\\
&=\int D[X_{l,[0,k]}]\;P\left( X_{j}^{(\nu)}(k+1)\big|X_{i,[0,k]}^{(\nu)},X_{j,[0,k]}^{(\nu)},X_{l,[0,k]}\right) \nonumber\\
&\qquad\qquad\qquad\;.\;P\left( X_{l,[0,k]}\big| X_{i,[0,k]}^{(\nu)},X_{j,[0,k]}^{(\nu)}\right) \label{eq:TE-conditionalexpression}
\end{align}
The first probability in the integral, which is the transition probability of $X_{j}$ in the full $d$-dimensional space, is analytically available. Additionally, similar to the procedure for the denominator in Eq. \ref{eq:temcestimate1}, samples from $P(X_{l,[0,k]}|X_{i,[0,k]}^{(\nu)},X_{j,[0,k]}^{(\nu)})$ are also available by sampling $X_{l}$ trajectories first from a $P_{0}(X_{l,[0,k]})$, and then applying the RR scheme. Thus the numerator in Eq. \ref{eq:temcestimate1} can also be evaluated in a Monte-Carlo fashion (see EM).

To summarize, for each of the $M_{1}$ pairs of $X_{i,[0,N]}$ and $X_{j,[0,N]}$ trajectories, we simulate a joint ensemble of $M_{2}$ new $X_{i}$ and $X_{l}$ trajectories to estimate the denominator in the logarithm in Eq. \ref{eq:temcestimate1}, and a separate ensemble of $M_{2}$ new $X_{l}$ trajectories to estimate the numerator (Fig. 1). The trajectories in each ensemble are resampled on-the-fly with the RR scheme, giving both the numerator and denominator in Eq. \ref{eq:temcestimate1} as Monte-Carlo averages. The computational cost thus scales as $2M_{1}M_{2}$. The transfer entropy estimate is unbiased\cite{reinhardt2023path} and the statistical accuracy can be arbitrarily improved by increasing $M_{1}$ and $M_{2}$. A pseudocode for the algorithm is available in SM-B. Aside from Schreiber's transfer entropy\cite{schreiber2000measuring}, other trajectory-based metrics of directional information transfer such as directed information\cite{massey1990causality}, conditional transfer entropy\cite{novelli2019large} and filtered transfer entropy\cite{chetrite2019information}, also can be derived from conditional distributions of trajectories. Hence TE-PWS can be used to compute all such metrics at similar cost, as shown in SM-G. 

\begin{figure}[t]
\centering
\includegraphics[width=8.6cm]{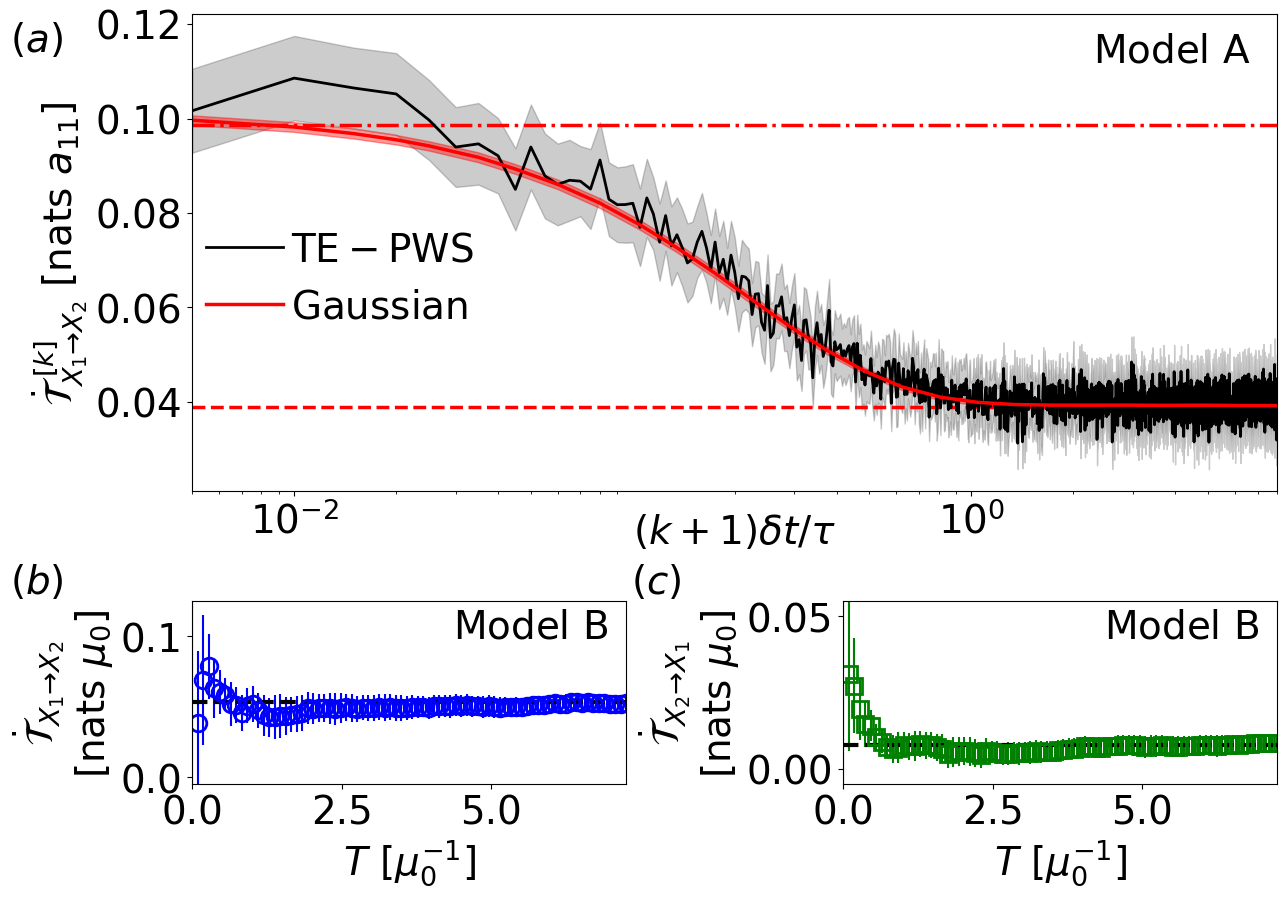}
\caption{Convergence of transfer entropy rate estimates in linear models. (a) shows $\dot{\mathcal{T}}_{X_{1}\to X_{2}}^{[k]}\equiv \mathcal{T}_{X_{1}\to X_{2}}^{[k]}/\delta t$ in the two-variable OU process (model A) with feedback as a function of the history length $k+1$. The black line is from TE-PWS, the solid red line is the exact result from the Gaussian framework, and the red dashed-dot and dashed lines are analytical results for $\dot{\mathcal{T}}_{X_{1}\to X_{2}}^{[k=0]}$ (single-step) and $\dot{\mathcal{T}}_{X_{1}\to X_{2}}$, respectively. (b) and (c) show transfer entropy rates in the linear stochastic gene expression model (model B)\cite{kiviet2014stochasticity}. Symbols are from TE-PWS and dashed lines are semi-analytical results.}
\label{fig:validation}
\end{figure}

We demonstrate the validity of the method by analyzing two examples of an Ornstein-Uhlenbeck (OU) process with feedback for which transfer entropy rates are exactly available since the trajectories are Gaussian-distributed. Eq. \ref{eq:langevin} represents an OU process when it is linear with $\mathbf{F}=-\mathbf{a}\mathbf{X}$, where $\mathbf{a}$ is a spring constant matrix. The first example is a two-variable OU process (model A) for which the transfer entropy rates are analytically available (see EM). Additionally, since the trajectories are Gaussian distributed, each term in Eq. \ref{eq:kdependentsum} can be exactly computed by estimating the covariance of trajectories, a commonly used approach called the Gaussian framework for the transfer entropy\cite{tostevin2009mutual,barnett2014mvgc} (see SM-H). 
On comparing these values with TE-PWS in Fig. \ref{fig:validation}a, we find that TE-PWS gives accurate and unbiased estimates of $\dot{\mathcal{T}}_{X_{1}\to X_{2}}^{[k]}\equiv \mathcal{T}_{X_{1}\to X_{2}}^{[k]}/\delta t$ for every $k$, converging to $\dot{\mathcal{T}}_{X_{1}\to X_{2}}$ for large $k$. 
A second benchmark is a three-variable model (model B) for describing the stochastic dynamics of gene expression and the growth rate of bacteria\cite{kiviet2014stochasticity}. The model is non-bipartite, meaning that the diffusion constant matrix is non-diagonal and the mutual information rates are not finite. Yet transfer entropy rates are finite and semi-analytically available\cite{chetrite2019information}. As shown in Fig. \ref{fig:validation}b and c, the steady-state transfer entropy rate from TE-PWS converges to the known values. Details about the models are provided in SM-A.

{\bf Data processing inequality.} We demonstrate the utility of TE-PWS by applying it to a three-node motif to study whether transfer entropies obey the Data Processing Inequality (DPI). Unidirectional flow of information between different nodes in a network leads to a DPI for the mutual information\cite{cover1999elements}. For a general three-variable process, if the flow of information is $X_{1}\to X_{2}\to X_{3}$, \textit{i.e.}, without feedback, $I(X_{1,[0,N]};X_{3,[0,N]}|X_{2,[0,N]})=0$; here the right arrows denote flow of information as mediated either via activation or repression. This leads by the chain rule to $I(X_{1,[0,N]};X_{3,[0,N]})\leq I(X_{1,[0,N]};X_{2,[0,N]})$\cite{cover1999elements}. As transfer entropy equals mutual information in the absence of feedback, it also obeys $\mathcal{T}_{X_{1}\to X_{3}}\leq\mathcal{T}_{X_{1}\to X_{2}}$.
This relation bounds the amount of information that can be transmitted from input to output through an intermediate variable, yet is only valid in the absence of feedback. In the presence of an $X_{2}\rightarrow X_{1}$ feedback, \textit{i.e.}, $X_{1}\rightleftarrows X_{2}\to X_{3}$, the mutual information continues to obey its DPI since $X_{1}$ and $X_{3}$ remain independent conditional on $X_{2}$ (see SM-I), while the transfer entropy formally doesn't\cite{james2016information,derpich2021directed}. This can be rationalized by considering the limit $X_{1}\leftarrow X_{2}\rightarrow X_{3}$, where $X_{2}$ controls both $X_{1}$ and $X_{3}$. Here the $X_{1}$ trajectories would still be predictive of fluctuations in $X_{3}$, thus $\mathcal{T}_{X_{1}\to X_{3}}\geq 0$ even as $\mathcal{T}_{X_{1}\to X_{2}}=0$. However, where the crossover from the feedforward to the feedback-dominated regime occurs, and to what extent transfer entropies can overcome the DPI in practice, is currently not understood. 

\begin{figure}[t]
\centering
\includegraphics[width=8.6cm]{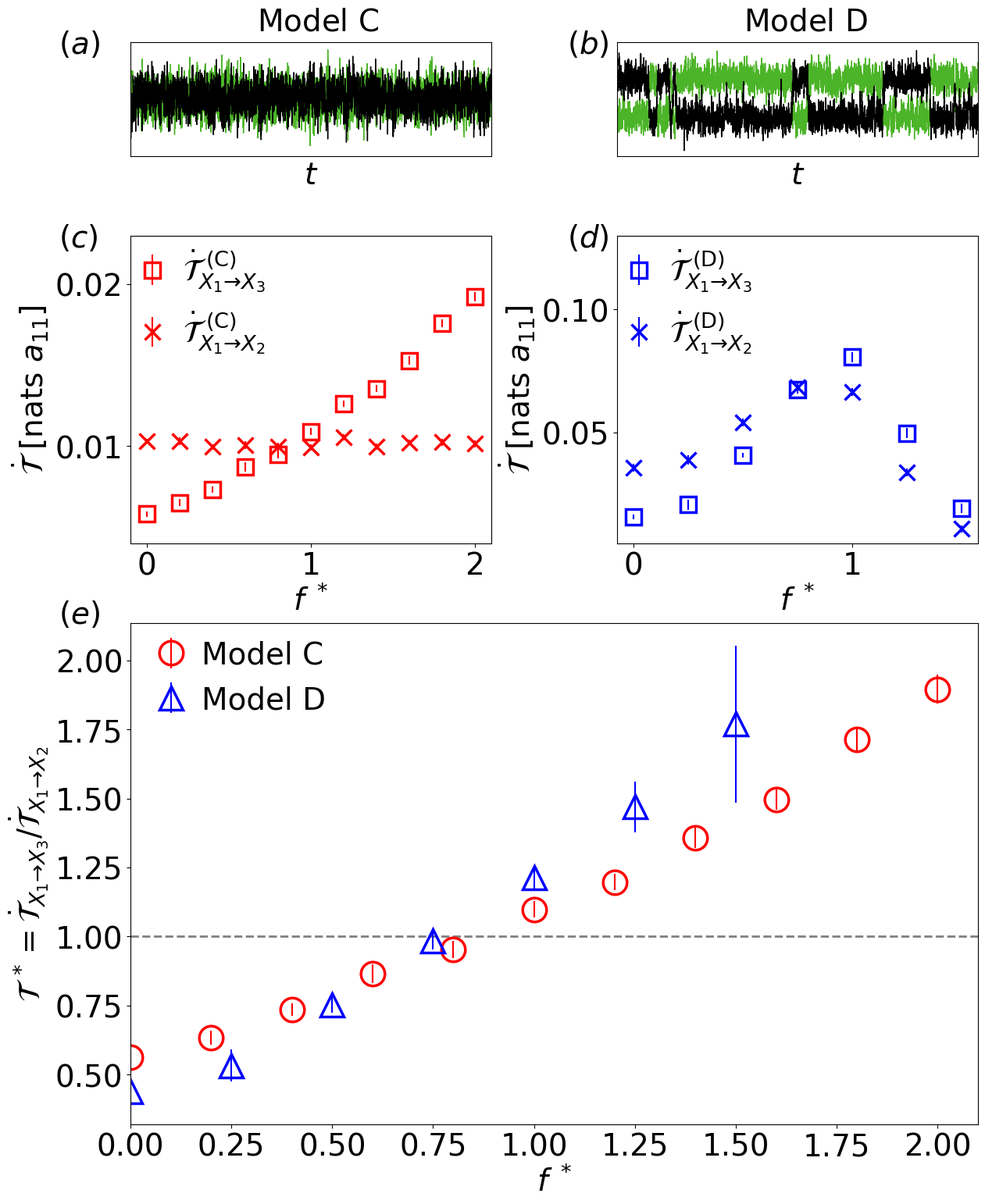}
\caption{(a) and (b) are typical $X_{1}$ (green) and $X_{2}$ (black) trajectories for models C and D at $a_{12}=2a_{21}=0.4$ and $a_{12}=a_{21}=-6$, respectively. (c) Transfer entropy rates in model C, denoted as $\dot{\mathcal{T}}_{X_{1}\to X_{3}}^{(\mathrm{C})}$ and $\dot{\mathcal{T}}_{X_{1}\to X_{2}}^{(\mathrm{C})}$, as a function of an increasing ratio of feedback to feedforward strength $f^{*}=a_{12}/a_{21}$, where the feedforward strength is kept constant at $a_{21}=0.2$. (d) Similar to (c) but for model D, where the feedforward strength is kept constant at $a_{21}=-4$. (e) Ratio of transfer entropy rates overcomes DPI bound (dashed line at $\mathcal{T}^{*}=1$) with increasing feedback.}
\label{fig:dpi}
\end{figure}

We implement the motif $X_{1}\rightleftarrows X_{2}\to X_{3}$ in two diffusive models of mutual repression between $X_{1}$ and $X_{2}$, labeled as models C and D. In both models, $X_{3}$ rapidly copies the state of $X_{2}$ such that the information loss from $X_{2}$ to $X_{3}$ is low. Model C is a three-dimensional OU process with linear feedback, where the ratio of the feedback to feedforward spring constants $f^{*}\equiv a_{12}/a_{21}$ is varied to study the violation of DPI. Model D is a nonlinear extension of C inspired by a genetic toggle switch\cite{warren2005chemical} where the drifts for $X_{1}$ and $X_{2}$ are changed to $F_{1}=-a_{11}X_{1}-a_{12}(1+X_{1}^{2})/(1+X_{1}^{2}+X_{2}^{2})$ and $F_{2}=-a_{22}X_{2}-a_{21}(1+X_{2}^{2})/(1+X_{1}^{2}+X_{2}^{2})$; we choose $a_{12}/a_{21}=f^{*}$ such that the two models can be compared. Other model parameters are provided in SM-A. Typical $X_{1}$ and $X_{2}$ trajectories at high feedback are shown in Figs. \ref{fig:dpi}a and b. Model C merely shows regression to the mean for both $X_{1}$ and $X_{2}$, while model D additionally shows switching between (low,high) and (high,low) values of $(X_{1},X_{2})$ in the regime of both strong feedforward and feedback coupling, around $f^{*}=1$.

We plot in Figs. \ref{fig:dpi}c and d transfer entropy rates for the two models. Each data point is obtained from steady state trajectories using TE-PWS. In model C, increasing the feedback strength $f^{*}$ keeps $\dot{\mathcal{T}}_{X_{1}\to X_{2}}$ unchanged while $\dot{\mathcal{T}}_{X_{1}\to X_{3}}$ monotonically increases. 
The former follows analytically from the expression for the transfer entropy in a two-dimensional OU process\cite{chetrite2019information} (see EM). The latter is an empirical result that arises because, when $f^{*}$ is increased, $X_{1}$ becomes increasingly correlated with $X_{2}$ while $X_{2}$ continues to be copied accurately by $X_{3}$. In contrast, in model D, both $\dot{\mathcal{T}}_{X_{1}\to X_{2}}$ and $\dot{\mathcal{T}}_{X_{1}\to X_{3}}$ peak near the switching regime, phenomenologically similar to how a bistable three-node motif with feedback has been shown to behave through approximate theory\cite{moor2023dynamic}. The values of the transfer entropy rates in model D are also amplified many-fold compared to model C due to stronger correlations between $X_{1}$ and $X_{2}$ resulting from the stronger, nonlinear coupling, with the concomitant switching. 

In Fig. \ref{fig:dpi}e we have plotted the degree of DPI violation in both models, quantified as the ratio of transfer entropy rates $\mathcal{T}^{*}\equiv\dot{\mathcal{T}}_{X_{1}\to X_{3}}/\dot{\mathcal{T}}_{X_{1}\to X_{2}}$, as a function of the ratio of the feedback to feedforward strengths, $f^{*}$. Surprisingly, we find that regardless of the nature of the variation of the individual transfer entropy rates with increasing feedback, \textit{i.e.}, monotonic or non-monotonic (Figs. \ref{fig:dpi}c and d), the ratio $\mathcal{T}^{*}$ monotonically increases with increasing feedback (Fig. \ref{fig:dpi} e). Moreover, the ratio overcomes the DPI bound when the strength of the feedback becomes comparable to that of the feedforward coupling. Our results thus show that when the feedback $X_{2}\to X_{1}$ dominates over the feedforward interaction, $X_{1}\to X_{2}$, the feedforward entropy $\mathcal{T}_{X_{1}\to X_{3}}$ becomes larger than the feedforward entropy $\mathcal{T}_{X_{1}\to X_{2}}$. The mutual information, on the other hand, continues to obey its DPI, $I(X_{1,[0,N]};X_{3,[0,N]})\leq I(X_{1,[0,N]};X_{2,[0,N]})$ (see SM-I). We expect further analytical work in the OU process to be able to support this empirical result.
Moreover, as the transfer entropy has been shown to limit the costs and benefits of information transmission\cite{mattingly2021escherichia,prokopenko2014transfer}, we expect the violation of the DPI to have significant physical consequences in functional networks.

{\bf Accuracy of approximate methods.}
An exact method is essential for the rigorous testing of the accuracy of approximate methods. This is now possible for the first time with our technique.
We have compared the ground-truth TE-PWS results with those obtained from the Gaussian framework and the KSG algorithm\cite{kraskov2004estimating}, in the linear model A and in the nonlinear model D (see Table \ref{tab:table1} in EM). We find that all approximate methods incur large systematic errors depending on the model. The Gaussian framework as expected gives accurate results in the linear model but fails in the nonlinear model. $\dot{\mathcal{T}}^{[k]}$ from the KSG method in the limit $(k\to\infty)$ has a systematic error due to a) its assumption of local uniformity of the distribution of the data points and b) the downsampling that becomes necessary to obtain a converged estimate (see SM-H). To avoid the systematic error, KSG is often truncated at only the $k=0$ term of Eq. \ref{eq:kdependentsum}, yielding the so-called single-step transfer entropy rate\cite{schreiber2000measuring,pahle2008information,lahiri2017information,chetrite2019information}. As the table shows and is demonstrated in Fig. \ref{fig:validation}a, the single-step transfer entropy rate is in general not an accurate estimate of the long-time limit. Interestingly, as the table shows, TE-PWS is not only exact but also highly cost efficient, being either comparable or orders of magnitude cheaper than approximate methods. This is because a) TE-PWS can calculate the transfer entropy for all $k$ in one run, and hence can average over all $k$ in the long time limit as shown in Eq. \ref{eq:transfer}, and b) for any given $k$, the cost for TE-PWS scales proportional to $k$, while for Gaussian and KSG the scaling is, respectively, $\sim k^{3}$ (see SM-H) and superlinear due to the cost of nearest neighbor search\cite{indyk1998approximate,ram2019revisiting} . 

In conclusion, we have developed a method that for the first time makes it possible to compute transfer entropies exactly for any stochastic model. 
We expect transfer entropies computed by TE-PWS to be used as \textit{ground truth} for a wide range of goals, such as the characterization and design of information flow in natural and engineered information processing systems, and causality detection.

{\bf Acknowledgement} We thank Manuel Reinhardt and Age Tjalma for useful discussions and Vahe Galstyan for a careful reading of the manuscript. This work is part of the Dutch Research Council (NWO) and was performed at the research institute AMOLF. This project has received funding from the European Research Council under the European Union’s Horizon 2020 research and innovation program (grant agreement No. 885065).

{\bf Data availability} Python implementations of TE-PWS for diffusive and jump processes, and data that reproduce the findings of this study, are openly available on Zenodo at \href {\doibase 10.5281/zenodo.13617365}{https://zenodo.org/doi/10.5281/zenodo.13617365}\cite{das_zenodo}.

\onecolumngrid
\appendix
\section{End Matter}
\twocolumngrid

\textbf{Analytical expression in two-dimensional OU process.} In a two-dimensional OU process $(X_{1},X_{2})$ with spring constants $a_{ij}$ and diffusion constant constants $D_{ij}$ for $i,j\in\{1,2\}$, the steady-state transfer entropy rate $\dot{\mathcal{T}}_{X_{1}\to X_{2}}$ is given by\cite{chetrite2019information}
\begin{align}\label{eq:te2d}
\dot{\mathcal{T}}_{X_{1}\to X_{2}}=\frac{1}{2}\left( r_{2}-a_{11}+\frac{D_{12}}{D_{22}}a_{21}\right)
\end{align}
where $r_{2}=[a_{11}^{2}+(D_{11}/D_{22})a_{21}^{2}-2(D_{12}/D_{22})a_{11}a_{21}]^{1/2}$. 
On the other hand, if Eq. \ref{eq:kdependentsum} is truncated at the first term after starting from a steady-state, the resultant approximation, called the single-step transfer entropy rate, is given by\cite{chetrite2019information}
\begin{align}\label{eq:te2d-1step}
\lim_{\delta t\to 0}\frac{1}{\delta t}\mathcal{T}_{X_{1}\to X_{2}}^{[0]}=\frac{1}{4D_{22}}\frac{a_{21}^{2}|\bf{V}|}{V_{22}}
\end{align}
where $\bf{V}$ is the stationary covariance matrix with elements $V_{ij}\equiv\langle X_{i}(0)X_{j}(0)\rangle$, analytical expressions for which can be found in \citen{chetrite2019information}.
In general, depending on parameter values such as feedback and noise strengths, Eqs. \ref{eq:te2d} and \ref{eq:te2d-1step} yield different values. Specifically, Eqs. \ref{eq:te2d} and \ref{eq:te2d-1step} have been used to plot the two distinct dashed lines in Fig. \ref{fig:validation}a. Additionally, we note that for model C in the section {\bf Data processing inequality} of the main text, the dynamics of $X_{1}$ and $X_{2}$ comprises that of a two-dimensional OU process because there is no feedback from $X_{3}$, only between $X_{1}$ and $X_{2}$. Hence the transfer entropy rate $\dot{\mathcal{T}}_{X_{1}\to X_{2}}$ is given by Eq. \ref{eq:te2d}, which does not depend on $X_{2}\to X_{1}$ feedback strength $a_{12}$, \textit{i.e.}, on $f^{*}$, as stated in the main text (see Fig.~\ref{fig:dpi}c).

{\bf Jump process.} For systems with jumps between a discrete number of states and finite waiting times between the jumps, such as a well-stirred chemical reaction network described by a master equation, or a neural spiking process, the dynamics is governed by a jump propensity matrix $\mathbf{Q}$ of dimensions $\mathcal{N}^{d}\times\mathcal{N}^{d}$, which describes jumps among the $\mathcal{N}$ states of each of the $d$ components of $\mathbf{X}$\cite{van1992stochastic}. In contrast with diffusive processes which must be time-discretized to simulate, jump processes can be simulated exactly with event-driven kinetic Monte-Carlo algorithms such as the Gillespie algorithm\cite{gillespie1976general}. Hence the transfer of information from $X_{i}$ to $X_{j}$ during the $(k+1)$-th trajectory segment occurs at all instants of time within the segment $X_{j,[k,k+1]}$, rather than at only the endpoint $X_{j}(k+1)$. Therefore in the definition of the stepwise increments to transfer entropy, $\mathcal{T}^{[k]}_{X_{i}\to X_{j}}$, in Eq. \ref{eq:transferentropy_probab}, the logarithms of the probabilities $P(X_{j}(k+1)|X_{i,[0,k]},X_{j,[0,k]})$ and $P(X_{j}(k+1)|X_{j,[0,k]})$ should be replaced with functionals of the entire $X_{i,[k,k+1]}$ and $X_{j,[k,k+1]}$ segments,
\begin{align}
\pi_{X_{i}\to X_{j}}&=-\int_{k\delta t}^{(k+1)\delta t}dt^{'}\lambda_{ij}(t^{'})+\sum_{\alpha=1}^{N_{j}}\ln\mathcal{Q}_{ij}(\alpha)\label{eq:marginal_propensity_1}\\
\pi_{X_{j}}&=-\int_{k\delta t}^{(k+1)\delta t}dt^{'}\lambda_{j}(t^{'})+\sum_{\alpha=1}^{N_{j}}\ln\mathcal{Q}_{j}(\alpha)\label{eq:marginal_propensity_2}\\
\mathcal{T}^{[k]}_{X_{i}\to X_{j}}&=\left\langle \pi_{X_{i}\to X_{j}}-\pi_{X_{j}}\right\rangle \label{eq:jumpte}
\end{align}
where $\alpha$ counts the jumps that change the state of $X_{j}$, $\lambda_{ij}$ and $\lambda_{j}$ are escape propensities for $X_{j}$ in the marginal spaces of $(X_{i},X_{j})$ and $(X_{j})$, respectively, and $\mathcal{Q}_{ij}(\alpha)$ and $\mathcal{Q}_{j}(\alpha)$ similarly are marginal jump propensities for the $\alpha$-th jump\cite{spinney2017transfer,moor2023dynamic}.
$Q_{ij}, \lambda_{ij}$ and $Q_{j},\lambda_{j}$ are obtained by marginalizing jump propensities from the full $\mathbf{X}$-space into the $(X_{i},X_{j})$ and $(X_{j})$ spaces, respectively\cite{spinney2017transfer,moor2023dynamic}.
The marginalization can be performed with TE-PWS in a Monte-Carlo fashion over conditional distributions of hidden variables, similar to the diffusive case (see SM-E). 
Prior work argues that the escape terms involving $\lambda_{ij}$ and $\lambda_{j}$ in Eqs. \ref{eq:marginal_propensity_1} and \ref{eq:marginal_propensity_2}, respectively, can be omitted since they cancel each other on average\cite{spinney2017transfer}. We show however in SM-F that the error in the transfer entropy estimate can be reduced by an order of magnitude by exploiting anti-correlated fluctuations between the escape and the jump terms.

\begin{table*}[t]
\caption{\label{tab:table1}Systematic errors and computational costs of approximate methods in the linear model A and the nonlinear model D. The input data size has been kept comparable whenever possible within non-prohibitive computational costs.
}
\begin{ruledtabular}
\begin{tabular}{ccccccc}
& &Exact&Gaussian&KSG $(k\to\infty)$
&KSG $(k=0)$\\ \hline
Linear model &Input data size $NM_{1}$& $1.6\times 10^{6}$&$1.6\times 10^{6}$&$1.6\times 10^{5}$&$1.6\times 10^{6}$\\
(model A) & $\dot{\mathcal{T}}_{1\to 2}~[10^{-2}~\mathrm{nats~} a_{11}]$& $3.98\pm0.06$& $3.90\pm0.33$& $5.82\pm0.07$& $8.98\pm3.39$\\
& CPU hours $\mathcal{C}$&$1.88$ & $1.35$&$75.92$ &$9.55$ \\ \hline
Nonlinear model &Input data size $NM_{1}$& $1.6\times 10^{6}$&$1.6\times 10^{6}$&$1.28\times 10^{5}$&$1.6\times 10^{6}$\\
(model D) & $\dot{\mathcal{T}}_{1\to 3}~[10^{-2}~\mathrm{nats~} a_{11}]$& $4.23\pm0.10$& $5.33\pm0.46$& $2.25\pm0.04$& $2.61\pm3.48$\\
& CPU hours $\mathcal{C}$&$1.90$ & $19.00$& $71.50$ &$9.13$ \\
\end{tabular}
\end{ruledtabular}
\end{table*}
\textbf{TE-PWS algorithm.} Central to TE-PWS is the computation of trajectory averages in a Monte-Carlo fashion over simulated trajectories, and the availability of trajectory probabilities on-the-fly in the full $d$-dimensional space. To implement TE-PWS for a given stochastic model, we need to specify the method to simulate the model and an explicit functional form of the trajectory probability.

The diffusion processes modeled by Eq. \ref{eq:langevin} can be simulated using an Euler-Maruyama scheme with a fixed small timestep. In our examples we have taken the timestep to be equal to the duration of the trajectory segments for implementing the RR scheme, $\delta t$, for convenience. The propagation equation for the $(k+1)$-th step is $\mathbf{X}(k+1)=\mathbf{X}(k)+\delta t\mathbf{F}(k)+\sqrt{\delta t}\bm{\psi}(k)$, where $\bm{\psi}$ is a Gaussian random vector with zero mean and variance $\langle\psi_{i}(k)\psi_{j}(k^{'})\rangle=2D_{ij}\delta_{k,k^{'}}$.
Making time discrete results in an $\mathcal{O}(\delta t)$ error which can be made arbitrarily small by systematically decreasing the timestep. The probability density of the change of state $\Delta \mathbf{X}(k)$ can be written analytically through Ito discretization of the Onsager-Machlup action as\cite{rogers2000diffusions,onsager1953fluctuations}
\begin{align}\label{eq:onsagermachlup}
P(\Delta\mathbf{X}(k))=&\frac{1}{(4\pi\delta t)^{d/2}|\mathbf{D}|}\exp\big[ -(\Delta\mathbf{X}(k)-\mathbf{F}(k)\delta t)^{T}\nonumber\\
&\qquad\cdot\mathbf{D}^{-1}(\Delta\mathbf{X}(k)-\mathbf{F}(k)\delta t)/4\delta t\big] 
\end{align}
where $|\mathbf{D}|$ is the determinant of the diffusion constant matrix. This form also holds for systems with inertia if the generalized coordinate vector $\mathbf{X}$ contains both positions and velocities\cite{das2019variational,lee2019thermodynamic}.

Jump processes can be simulated with a Gillespie algorithm, which is exact, \textit{i.e.}, does not make a timestep error\cite{gillespie1976general}. 
The probability density of a trajectory segment $\mathbf{X}_{[t,t+\delta t]}$ is written as
\begin{align}\label{eq:gillespie}
\ln P(\mathbf{X}_{[t,t+\delta t]})=-\int_{t}^{t+\delta t}dt^{'}\lambda(t^{'})+\sum_{\alpha=1}^{N_{\mathrm{tot}}}\ln\mathcal{Q}_{\alpha}(t_{\alpha})
\end{align}
where $\lambda(t^{'})$ is the escape propensity from state $\mathbf{X}(t^{'})$, given by the sum of the jump propensities $\mathcal{Q}_{\beta}(t^{'})$ taking the system out of the state $\mathbf{X}(t^{'})$ at time $t^{'}$, and $\alpha$ sums over all jumps in the full trajectory, $N_{\mathrm{tot}}$ in number, occurring at times $t_{\alpha}$.

The elementary steps of the TE-PWS algorithm are the same for both diffusive and jump processes. In brief:
\begin{enumerate}
    \item Propagate $M_{1}$ trajectories of $X_{i}$ and $X_{j}$ in the full $d$-dimensional space. 
    These trajectories are henceforth labeled with $(\nu)$.

\makeatletter
\newenvironment{fullwidth}
    {\par
     \setlength{\@totalleftmargin}{0pt}%
     \setlength{\linewidth}{\hsize}%
     \list{}{\setlength{\leftmargin}{0pt}}
     \item\relax}
    {\endlist}
\makeatother
    \begin{fullwidth} Steps 2-5 are for computing the numerator in Eq. \ref{eq:temcestimate1} for diffusive processes and the first term in Eq. \ref{eq:jumpte} for jump processes. \end{fullwidth}
    
    \item For each pair of $(X_{i}^{(\nu)},X_{j}^{(\nu)})$ trajectories, propagate $M_{2}$ trajectories of hidden variables $X_{l}$ using the chosen reference distribution $P_{0}(X_{l,[0,k]})$. Initial conditions should be sampled from the same joint distribution $P(X_{i}(0),X_{j}(0),X_{l}(0))$ that $X_{i}^{(\nu)}(0)$ and $X_{j}^{(\nu)}(0)$ were drawn from.
    \item After every $\delta t$ time, recalculate logarithmic weights $w^{(\mu)}$ for the trajectories defined as the logarithm of the ratio between the joint distribution $P(X_{i,[0,k]}^{(\nu)},X_{j,[0,k]}^{(\nu)},X_{l,[0,k]})$ and the reference distribution $P_{0}(X_{l,[0,k]})$, using Eqs. \ref{eq:onsagermachlup} and \ref{eq:gillespie}. This ratio is proportional to the ratio between the conditional distribution $P(X_{l,[0,k]}|X_{i,[0,k]}^{(\nu)},X_{j,[0,k]}^{(\nu)})$ and the reference distribution $P_{0}(X_{l,[0,k]})$, as proven in SM-C. Then calculate the uniformity in the weights with a uniformity parameter $\kappa=(\sum_{\mu} \exp w^{(\mu)})^{2}/\sum_{\mu}\exp (2w^{(\mu)})$, where sums of exponentials of weights are always performed with the Log-Sum-Exp trick\cite{Gundersen_2020}.
    \item Calculate the contribution to the transfer entropy in the space of $(X_{i},X_{j})$ at the $(k+1)$-th step, denoted as $\mathcal{T}_{a}^{(\nu)}[k]$. This is done by computing the average in Eq. \ref{eq:TE-conditionalexpression} for diffusive processes, and taking an expectation of Eq. \ref{eq:marginal_propensity_1} for jump processes (see SM-E).
    \item If $\kappa<M_{2}/2$, resample the $M_{2}$ trajectories with the accumulated weights $w^{(\mu)}$ (see Fig. \ref{fig:schematic}c). This means we sample $M_{2}$ trajectories randomly with weights $w^{(\mu)}$ from the simulated trajectory ensemble, with replacement. Set all weights to zero after every resampling.

    \begin{fullwidth} Step 6 is for computing the denominator in Eq. \ref{eq:temcestimate1} for diffusive processes and the second term in Eq. \ref{eq:jumpte} for jump processes. \end{fullwidth}
        
    \item Akin to steps 2-5, for each $(X_{j}^{(\nu)})$ trajectory, propagate $M_{2}$ trajectories of $(X_{i},X_{l})$ using reference distribution $P_{0}(X_{i,[0,k]},X_{l,[0,k]})$; accumulate weights for each trajectory as the ratio between the joint distribution $P(X_{i,[0,k]},X_{j,[0,k]}^{(\nu)},X_{l,[0,k]})$ and the reference distribution $P_{0}(X_{i,[0,k]},X_{l,[0,k]})$; compute the contribution to the transfer entropy in the $(X_{j})$ space at the $(k+1)$-th step, denoted as $\mathcal{T}_{b}^{(\nu)}[k]$ (Eqs. \ref{eq:temcestimate2} and \ref{eq:marginal_propensity_2}); and resample if the uniformity in the weights is low, \textit{i.e.}, $\kappa<M_{2}/2$.
    
    \item Finally, compute $\mathcal{T}_{X_{i}\to X_{j}}$ by combining $\mathcal{T}_{a}^{(\nu)}$ and $\mathcal{T}_{b}^{(\nu)}$ from all timesteps using Eqs. \ref{eq:temcestimate1} and \ref{eq:jumpte} for diffusive and jump processes, respectively.
\end{enumerate}
For clarity, a pseudocode to compute the transfer entropy $\mathcal{T}_{X_{1}\to X_{3}}$ in a three-variable process is given in SM-B. We note that TE-PWS is easily parallelized as it is a Monte-Carlo algorithm. Specifically, steps 1-7 can be executed independently for each of the $M_{1}$ trajectories mentioned in step 1, using parallel processors. The transfer entropy is computed at the end as an average over the results from all processors.

\onecolumngrid
\clearpage
\setcitestyle{super}

\newcommand{\beginsupplement}{%
        \setcounter{table}{0}
        \renewcommand{\theequation}{S.\arabic{equation}}
        \renewcommand{\thetable}{S\arabic{table}}%
        \renewcommand{\thefigure}{S\arabic{figure}}%
}
\renewcommand{\citenumfont}[1]{S#1} 
\renewcommand{\bibnumfmt}[1]{[S#1]} 

\beginsupplement
\title{Supplemental Material: \\Exact computation of Transfer Entropy with Path Weight Sampling}


\date{\today}

\maketitle

\thispagestyle{empty}
\onecolumngrid



\setcounter{equation}{0}
\setcounter{figure}{0}
\setcounter{table}{0}
\makeatletter
\renewcommand{\theequation}{S\arabic{equation}}
\renewcommand{\thefigure}{S\arabic{figure}}
\section{SM-A: Simulation details}
\textbf{Model A.} For Fig. 2a in the main text, the model (model A) is a two-dimensional OU process with parameters $a_{11}=a_{22}=1$, $a_{12}=a_{21}=0.9$, $D_{11}=0.2$, $D_{22}=1$ and $D_{12}=0$. Additionally, $\delta t=0.01 a_{11}^{-1}$, $M_{1}=10^{6}$ and $M_{2}=100$ for TE-PWS. For the Gaussian framework, our protocol for obtaining the converged transfer entropy rates in the infinite data limit are described in the section \textbf{SM-H: Comparison with approximate methods}.

\textbf{Model B.} Figs. 2b and c in the main text show transfer entropies in the minimal stochastic model for gene expression from Refs. \cite{smkiviet2014stochasticity,smlahiri2017information,smchetrite2019information}. The model (model B) is a three-dimensional OU process with spring constants $a_{11}=\mu_{E}+\mu_{0}T_{\mu E}(T_{EG}-1)$, $a_{12}=-\mu_{0}(T_{EG}-1)$, $a_{13}=\mu_{0}T_{EG}$, $a_{21}=-T_{\mu E}[\beta_{G}-\mu_{E}-\mu_{0}T_{\mu E}(T_{EG}-1)]$, $a_{22}=-[\mu_{0}T_{\mu E}(T_{EG}-1)-\beta_{G}]$, $a_{23}=-[\beta_{G}-\beta_{\mu}-\mu_{0}T_{\mu E}T_{EG}]$, $a_{31}=a_{32}=0$ and $a_{33}=\beta_{\mu}$, and diffusion constants $D_{11}=D_{E}\mu_{0}^{2}$, $D_{22}=\beta_{\mu}\eta_{\mu}^{2}+\beta_{G}\eta_{G}^{2}+D_{E}\mu_{0}^{2}T_{\mu E}^{2}$, $D_{33}=\beta_{\mu}\eta_{\mu}^{2}$, $D_{12}=D_{21}=D_{E}\mu_{0}^{2}T_{\mu E}$, $D_{13}=D_{31}=0$ and $D_{23}=D_{32}=\beta_{\mu}\eta_{\mu}^{2}$, where the experimentally determined values of the parameters are $\mu_{0}=0.23 h^{-1}$, $\beta_{\mu}=\beta_{G}=0.33 h^{-1}$, $\beta_{E}=5.63 h^-1$, $\eta_{E}=1.03$, $\eta_{\mu}=0.16$, $\eta_{G}=0.22$, $T_{EE}=1$, $T_{\mu E}=0.7$, $T_{EG}=1.3$, $\mu_{E}=\mu_{0}(1+T_{\mu E}-T_{EE})$ and $D_{E}=\eta_{E}^{2}/\beta_{E}$.\cite{smkiviet2014stochasticity,smchetrite2019information}. We also used $\delta t=9.2\times10^{-5}\mu_{0}^{-1}$, $M_{1}=96$ and $M_{2}=200$.

\textbf{Models C and D.} For models C and D in Fig. 3 in the main text, we have chosen $a_{11}=a_{22}=1$, $a_{33}=-a_{32}=2$, $a_{13}=a_{31}=a_{23}=0$ and $D_{ij}=\delta_{ij}$. Additionally, $\delta t=0.01 a_{11}^{-1}$, $M_{1}=100$ and $M_{2}=1000$.

In all simulations, trajectories have been initialized from a collection of the corresponding steady-state distributions, including for marginalization in TE-PWS. For details of initializing TE-PWS trajectories, see the SM section \textbf{SM-D: Choice of reference distribution}. Error bars for all results are computed as twice the standard deviations using multiple sets of statistically independent simulations. For comparison with the performance of TE-PWS, the KSG method with 4 nearest neighbors (the `first' KSG algorithm in \citen{smkraskov2004estimating}) was implemented with the openly available JIDT toolkit\cite{smlizier2014jidt}. The computational cost of all techniques was monitored, for a total of 40 independent samples for each technique, with the openly available process analysis workbench Procpath\cite{smprocpath}. The CPU cost mentioned in Table 1 of the main text corresponds to a calculation for only the largest studied history length $(k+1)$ in each case. For details on the convergence of the transfer entropy rate from Gaussian and the KSG algorithms (data presented in Table 1 of the main text) with respect to the history length $(k+1)$ and the data size $NM_{1}$, see the SM section \textbf{SM-H: Comparison with approximate methods}.

\section{SM-B: Pseudocode}
A pseudocode for computing the transfer entropy $\mathcal{T}_{X_{1}\to X_{3}}$ in a three-variable stochastic process via the TE-PWS algorithm is provided in Algorithm \ref{algo:tepws}.
\begin{figure*}
\begin{minipage}{\linewidth}
\begin{algorithm}[H]
	\caption{Transfer Entropy-Path Weight Sampling (TE-PWS)}\label{algo:tepws}
	\begin{algorithmic}[1]
        \State \textbf{objective} computation of the transfer entropy $\mathcal{T}_{X_{1}\to X_{3}}$ in a three-variable stochastic process.
		\State \textbf{inputs} model parameters for simulating $\mathbf{X}(t)$
		\State \textbf{parameters} timestep $\delta t$; number of timesteps $N$; number of trajectories for Monte-Carlo averages $M_{1}$ and $M_{2}$; resampling indicator $\kappa$
		\State \textbf{initialize} Define trajectory labels $\nu$ and $\mu$; timestep variable $k$; cumulative transfer entropy array $\mathcal{T}[0:N]$; two terms (factors) of transfer entropy  $\mathcal{T}_{a}^{(\nu)}[0:N]$ and $\mathcal{T}_{b}^{(\nu)}[0:N]$ respectively. 
		\State $\nu\gets0$
        \State $\mathcal{T}[0:N]=0$
		\Repeat
        \State Generate $M_{1}$ trajectories $(X_{1,[0,N]}^{(\nu)}, X_{2,[0,N]}^{(\nu)},X_{3,[0,N]}^{(\nu)})$ jointly.
        \State $k\gets 0$ 
        \Comment{Accessing $P(X_{2,[0,N]}|X_{1,[0,N]},X_{3,[0,N]})$}
        \State $\kappa\gets M_{2}$
        \State \textbf{initialize} Generate $M_{2}$ samples of initial conditions $X_{2}^{(\mu)}(0)$ labeled by $\mu$ from a steady-state trajectory; weights in log scale $w^{(\mu)}=0$.
        \Repeat
		\If {$\kappa<M_{2}/2$}
		    \State Resample $M_{2}$ configurations from $X_{2}^{(\mu)}(k)$ with weights $\exp w^{(\mu)}$.
            \State $w^{(\mu)}\gets 0$ for all $\mu$
		\EndIf        
        \State Propagate reference dynamics $X_{2,[k,k+1]}^{(\mu)}$ as samples from $P_{0}(X_{2,[k,k+1]}|X_{2,[0,k]})$.
        \State $w^{(\mu)}\gets w^{(\mu)}+\ln P\left( X_{1,[k,k+1]}^{(\nu)},X_{2,[k,k+1]}^{(\mu)},X_{3,[k,k+1]}^{(\nu)}\big| X_{1,[0,k]}^{(\nu)},X_{2,[0,k]}^{(\mu)},X_{3,[0,k]}^{(\nu)}\right) -\ln P_{0}\left( X_{2,[k,k+1]}^{(\mu)}\big| X_{2,[0,k]}^{(\mu)}\right) $
        \State $\kappa\gets (\sum_{\mu} \exp w^{(\mu)})^{2}/\sum_{\mu}\exp (2w^{(\mu)})$
        \State Compute $\mathcal{T}_{a}^{(\nu)}[k]$ using Eq. 9 or 12 of the main text for a diffusion or jump process respectively.
        \State $k\gets k+1$
        \Until{$k=N$}
        \State $k\gets 0$ 
        \Comment{Accessing $P(X_{1,[0,N]},X_{2,[0,N]}|X_{3,[0,N]})$}
        \State $\kappa\gets M_{2}$
        \State \textbf{initialize} Generate $M_{2}$ samples of initial conditions $(X_{1}^{(\mu)}(0),X_{2}^{(\mu)}(0))$ labeled by $\mu$ from a steady-state trajectory; weights in log scale $w^{(\mu)}=0$.
        \Repeat
		\If {$\kappa<M_{2}/2$}
		    \State Resample $M_{2}$ configurations from $(X_{1}^{(\mu)}(k),X_{2}^{(\mu)}(k))$ with weights $\exp w^{(\mu)}$. 
            \State $w^{(\mu)}\gets 0$ for all $\mu$
		\EndIf        
        \State Propagate reference dynamics $(X_{1,[k,k+1]}^{(\mu)},X_{2,[k,k+1]}^{(\mu)})$ as samples from $P_{0}(X_{1,[k,k+1]},X_{2,[k,k+1]}|X_{1,[0,k]},X_{2,[0,k]})$.
        \State $w^{(\mu)}\gets w^{(\mu)}+\ln P\left( X_{1,[k,k+1]}^{(\mu)},X_{2,[k,k+1]}^{(\mu)},X_{3,[k,k+1]}^{(\nu)}\big| X_{1,[0,k]}^{(\mu)},X_{2,[0,k]}^{(\mu)},X_{3,[0,k]}^{(\nu)}\right) $
        \State $w^{(\mu)}\gets w^{(\mu)}-\ln P_{0}\left( X_{1,[k,k+1]}^{(\mu)},X_{2,[k,k+1]}^{(\mu)}\big| X_{1,[0,k]}^{(\mu)},X_{2,[0,k]}^{(\mu)}\right) $
        \State $\kappa\gets (\sum_{\mu} \exp w^{(\mu)})^{2}/\sum_{\mu}\exp (2w^{(\mu)})$
        \State Compute $\mathcal{T}_{b}^{(\nu)}[k]$ using Eq. 8 or 13 in the main text for a diffusion or jump process respectively.
        \State $k\gets k+1$
        \Until{$k=N$}
        \State $\mathcal{T}[0:N]\gets\mathcal{T}[0:N]+\mathcal{T}^{(\nu)}_{a}[0:N]-\mathcal{T}^{(\nu)}_{b}[0:N]$
		\State $\nu\gets \nu+1$
		\Until{$\nu=M_{1}$}
        \State $\mathcal{T}[0:N]\gets \mathcal{T}[0:N]/M_{1}$
	\end{algorithmic}
\end{algorithm}
\end{minipage}
\end{figure*}

\section{SM-C: Proof of the efficiency of the RR scheme}
Here we prove that the Rosenbluth-Rosenbluth (RR) scheme in TE-PWS achieves optimal efficiency for a Monte-Carlo evaluation of the transfer entropy. For clarity, we start by reproducing Eq. 8 of the main text here, representing a typical marginalization procedure in TE-PWS,
\begin{align}
P\left( X_{j,[0,k]}^{(\nu)}\right) &=\frac{1}{M_{2}}\sum_{\mu} \frac{P\left( X_{i,[0,k]}^{(\mu)},X_{j,[0,k]}^{(\nu)},X_{l,[0,k]}^{(\mu)}\right) }{P_{0}\left( X_{i,[0,k]}^{(\mu)},X_{l,[0,k]}^{(\mu)}\right) }\label{eq:sm_temcestimate2}
\end{align}
TE-PWS performs this average with high statistical efficiency by preferentially sampling rare large values of the summand. Similar to PWS as discussed in \citen{smreinhardt2023path}, this is implemented using the RR scheme, in which the ensemble of $M_{2}$ trajectories is resampled after every $\delta t$ time (see Fig. 1 of main text). For example, in order to compute $P(X_{j,[0,N]}^{(\nu)})$, an ensemble of $X_{i}$ and $X_{l}$ trajectories is simulated. At the $(k+1)$-th step, the resampling weight used for the $\mu$-th trajectory in the ensemble is
\begin{align}
&\widehat{g}\big[ X_{l,[0,k+1]}^{(\mu)}\big] =  P\left( X_{i,[k,k+1]}^{(\mu)},X_{j,[k,k+1]}^{(\nu)},X_{l,[k,k+1]}^{(\mu)}\Big| X_{i,[0,k]}^{(\mu)},X_{j,[0,k]}^{(\nu)},X_{l,[0,k]}^{(\mu)}\right) /\;P_{0}\left( X_{i,[k,k+1]}^{(\mu)},X_{l,[k,k+1]}^{(\mu)}\Big| X_{i,[0,k]}^{(\mu)},X_{l,[0,k]}^{(\mu)}\right) \label{eq:ghat}
\end{align}
which is analytically available.
We show below that iteratively resampling the newly generated trajectories with this weight changes the trajectory distribution optimally such that the summand in Eq. \ref{eq:sm_temcestimate2} is a constant, \textit{i.e.}, has a variance of zero, achieving perfect sampling. We will show this by first showing that if after the $k$-th step the trajectories have an optimal distribution, they stay optimal after the $(k+1)$-th step. Then, combined with the fact that the distribution of initial conditions is by construction optimal, we will conclude by induction that the distribution stays uniform during the entire duration of the trajectory.

The optimal choice for $P_{0}$ would be the hypothetical $P_{0}(X_{i,[0,N]},X_{l,[0,N]})=P(X_{i,[0,N]},X_{l,[0,N]}|X_{j,[0,N]}^{(\nu)})$ as it makes the summand in Eq. \ref{eq:sm_temcestimate2} a constant independent of the index $\mu$, resulting in a zero-variance estimate of $P(X_{j,[0,N]}^{(\nu)})$. The purpose of the RR scheme is to bias the simulated distribution $P_{0}(X_{i,[0,N]},X_{l,[0,N]})$ towards the optimal distribution $P(X_{i,[0,N]},X_{l,[0,N]}|X_{j,[0,N]}^{(\nu)})$. For the inductive argument, assume that after the $k$-th step, the trajectories are distributed optimally according to $P(X_{i,[0,k]},X_{l,[0,k]}|X_{j,[0,k]}^{(\nu)})$. After the next propagation step, the probability of each trajectory changes to a product of its previous value and the probability of the new segment,
\begin{align}
w^{(\mu)}&=P\left( X_{i,[0,k]}^{(\mu)},X_{l,[0,k]}^{(\mu)}\Big|X_{j,[0,k]}^{(\nu)}\right) .\;P_{0}\left( X_{i,[k,k+1]}^{(\mu)},X_{l,[k,k+1]}^{(\mu)}\Big| X_{i,[0,k]}^{(\mu)},X_{l,[0,k]}^{(\mu)}\right) \nonumber\\
&=P\left( X_{i,[0,k]}^{(\mu)},X_{j,[0,k]}^{(\nu)},X_{l,[0,k]}^{(\mu)}\right) .\;P_{0}\left( X_{i,[k,k+1]}^{(\mu)},X_{l,[k,k+1]}^{(\mu)}\Big| X_{i,[0,k]}^{(\mu)},X_{l,[0,k]}^{(\mu)}\right) \big/\;P\left( X_{j,[0,k]}^{(\nu)}\right) \label{eq:propagatedw} 
\end{align}
Then we resample the trajectories with weight $\widehat{g}$ given in Eq. \ref{eq:ghat}. The probability distribution of these trajectories after this resampling step  becomes proportional to the product of Eqs. \ref{eq:ghat} and \ref{eq:propagatedw},
\begin{align}\label{eq:productweight}
w^{(\mu)}\widehat{g}\big[ X_{l,[0,k+1]}^{(\mu)}\big] &=\frac{P\left( X_{i,[0,k+1]}^{(\mu)},X_{j,[0,k+1]}^{(\nu)},X_{l,[0,k+1]}^{(\mu)}\right) }{P\left( X_{j,[0,k]}^{(\nu)}\right) }
\end{align}
where Bayes' theorem has been used to condense the numerator.
The normalization constant for this probability is obtained by summing Eq. \ref{eq:productweight} over all $X_{i,[0,k+1]}^{(\mu)}$ and $X_{l,[0,k+1]}^{(\mu)}$ trajectories, which gives $P\left( X_{j,[0,k+1]}^{(\nu)}\right) \big/ P\left( X_{j,[0,k]}^{(\nu)}\right) $. Dividing Eq. \ref{eq:productweight} by this normalization constant gives the new normalized probability distribution as $P(X_{i,[0,k+1]}^{(\mu)},X_{l,[0,k+1]}^{(\mu)}|X_{j,[0,k+1]}^{(\nu)})$. Hence the trajectories remain distributed optimally after resampling. 

To complete the proof by induction, we also need to show that the initial conditions of the $X_{i,[0,k]}^{(\mu)}$ and $X_{l,[0,k]}^{(\mu)}$ trajectories are consistent with the optimal distribution $P(X_{i,[0,k]},X_{l,[0,k]}|X_{j,[0,k]}^{(\nu)})$. This requirement is met by construction, because the initial conditions for the $X_{i}^{(\mu)}$ and $X_{l}^{(\mu)}$ trajectories are sampled from the same joint distribution $P(\mathbf{X}(0))$ as that from which $X_{j}^{(\nu)}(0)$ is drawn; if one is interested in the steady-state transfer entropy rates, this distribution is the steady-state joint distribution where $X_{j}=X_{j}^{(\nu)}(0)$ is given. This ingredient, together with the fact that the resampling procedure preserves the optimal distribution as described above, guarantees that, by induction, the RR scheme generates samples from the optimal conditional distribution $P(X_{i,[0,k]},X_{l,[0,k]}|X_{j,[0,k]}^{(\nu)})$ at every step. Resampling is algorithmically performed with a stratified resampling technique which is computationally efficient\cite{smdouc2005comparison}. Samples from the optimal distribution are then used in Eq. \ref{eq:sm_temcestimate2} to compute the denominator in Eq. 6 of the main text, reproduced here,
\begin{align}\label{eq:sm_temcestimate1}
\mathcal{T}_{X_{i}\to X_{j}}&=\frac{1}{M_{1}}\sum_{\nu}\sum_{k}\ln\frac{P\left( X^{(\nu)}_{j}(k+1)\big| X^{(\nu)}_{i,[0,k]},X^{(\nu)}_{j,[0,k]}\right) }{P\left( X^{(\nu)}_{j}(k+1)\big| X^{(\nu)}_{j,[0,k]}\right) }
\end{align}
The numerator is obtained similarly by generating samples from the corresponding conditional distribution $P(X_{l,[0,k]}|X_{i,[0,k]}^{(\nu)},X_{j,[0,k]}^{(\nu)})$ with the RR scheme. Conditional distributions of trajectories of all combinations of variables can be similarly sampled by repeating this procedure.

\section{SM-D: Choice of reference distribution}
The transfer entropy estimated from TE-PWS is exact for any choice of the reference probability $P_{0}$ because of the RR scheme. However, the number of times resampling needs to be performed depends on how large the variance of the summand in Eq. \ref{eq:sm_temcestimate2} is. A better choice of $P_{0}$ results in a smaller variance at the same computational cost. As discussed earlier, for computing $P(X_{j,[0,N]}^{(\nu)})$ for example, the ideal choice for $P_{0}$ would be $P_{0}(X_{i,[0,N]},X_{l,[0,N]})=P(X_{i,[0,N]},X_{l,[0,N]}|X_{j,[0,N]}^{(\nu)})$, which is not known \textit{a priori} and is impossible to directly sample from. We therefore choose a distribution $P_{0}(X_{i,[0,N]},X_{l,[0,N]})$ that uses the past trajectory of $X_{j,[0,N]}^{(\nu)}$ at every timestep to compute the drift and diffusion terms for $X_{i}$ and $X_{l}$, similar to the original dynamics of the system in the full $d$-dimensional space. This keeps the reference distribution close to the target conditional distribution while being analytically known and easy to sample from.

As an example, consider the three-dimensional OU process discussed before (model B), $\dot{\mathbf{X}}=-\mathbf{a}\mathbf{X}+\bm{\xi}$ where $\mathbf{a}$ is the spring-constant matrix and $\bm{\xi}$ is a Gaussian white noise with zero mean and covariance matrix $2\mathbf{D}$. The diffusion constant matrix $\mathbf{D}$ may have nonzero off-diagonal elements. In the full three-dimensional space, propagating the natural dynamics involves computing the drifts and sampling three correlated noise components from the joint Gaussian distribution, which we will call $\mathbb{G}(\xi_{i},\xi_{j},\xi_{l})$. For sampling trajectories $X_{i,[0,N]}^{(\mu)}$ and $X_{l,[0,N]}^{(\mu)}$, for a given $X_{j,[0,N]}^{(\nu)}$, from a distribution $P_{0}(X_{i,[0,N]}^{(\mu)},X_{l,[0,N]}^{(\mu)})$ that is as close as possible to $P(X_{i,[0,N]}^{(\mu)},X_{l,[0,N]}^{(\mu)}|X_{j,[0,N]}^{(\nu)})$, we first initialize the trajectories from the single-time conditional distribution $P(X_{i},X_{l}|X_{j}^{(\nu)}(0))$, by either storing a representative list of steady-state $X_{i}$ and $X_{l}$ configurations such that the corresponding $X_{j}$ is within a given bin-width of $X_{j}^{(\nu)}(0)$, or by explicitly constructing a three-dimensional histogram for the steady-state concentration of all three variables with a given bin-width. The first approach was used for Fig. 2a of the main text, where obtaining $\dot{\mathcal{T}}^{[k]}_{X_{1}\to X_{2}}$ accurately for every value of $k$ was crucial, including for small values of $k$. If only the transfer entropy rate in the long-time limit is required, the second method is sufficient and has been used for all other TE-PWS simulations in the paper.

We then propagate $X_{i}^{(\mu)}$ and $X_{l}^{(\mu)}$ trajectories with the equations of motion 
\begin{align}
\dot{X}_{i}^{(\mu)}(k)&=-a_{11}X_{i}^{(\mu)}(k)-a_{12}X^{(\nu)}_{j}(k)-a_{13}X_{l}^{(\mu)}(k)+\tilde{\xi}_{i}^{(\mu)}(k)\\
\dot{X}_{l}^{(\mu)}(k)&=-a_{31}X_{i}^{(\mu)}(k)-a_{32}X^{(\nu)}_{j}(k)-a_{33}X_{l}^{(\mu)}(k)+\tilde{\xi}_{l}^{(\mu)}(k)
\end{align}
where $\tilde{\xi}_{i}^{(\mu)}$ and $\tilde{\xi}_{l}^{(\mu)}$ are Gaussian white noises whose distributions should be commensurate with the existing noises $\xi_{j}^{(\nu)}$ in the $X_{j,[0,N]}^{(\nu)}$ trajectory. However, given only the $X_{j,[0,N]}^{(\nu)}$ trajectory, the noises $\xi_{j}^{(\nu)}(k)$ are not uniquely known since the drifts $F_{j}^{(\nu)}(k)$ are unknown; hence, the distributions of $\tilde{\xi}_{i}^{(\mu)}$ and $\tilde{\xi}_{l}^{(\mu)}$ are not yet uniquely defined. To fully specify these distributions, we now make a choice for deriving approximate noises in $X_{j,[0,N]}^{(\nu)}$, called $\tilde{\xi}_{j}^{(\nu\mu)}$, by assuming that the drift in $X_{j}$ depends, besides $X_{j}^{(\nu)}(k)$ itself, on  $X_i^{(\mu)}(k)$ and $X_l^{(\mu)}(k)$, \textit{i.e.} the drift is taken to be $-a_{21}X_{i}^{(\mu)}(k)-a_{22}X^{(\nu)}_{j}(k)-a_{23}X_{l}^{(\mu)}(k)$. Next, obtaining $\dot{X}_{j}^{(\nu)}(k)$ from the given $X_{j}^{(\nu)}$ trajectory and with the drift now specified, we solve, at every timestep,
\begin{align}
\dot{X}_{j}^{(\nu)}(k)&=-a_{21}X_{i}^{(\mu)}(k)-a_{22}X^{(\nu)}_{j}(k)-a_{23}X_{l}^{(\mu)}(k)+\tilde{\xi}_{j}^{(\nu\mu)}(k)
\end{align}
for $\tilde{\xi}_{j}^{(\nu\mu)}(k)$, which can be done by simply transposing the equation even if the original dynamics was not linear. Now given $\tilde{\xi}_{j}^{(\nu\mu)}(k)$, from which distribution should we generate $\tilde{\xi}_{i}^{(\mu)}(k)$ and $\tilde{\xi}_{l}^{(\mu)}(k)$? Here we use the conditional noise distribution in the natural dynamics of the system in order to keep the reference distribution close to optimal. Given the multivariate Gaussian distribution $\mathbb{G}(\xi_{i},\xi_{j},\xi_{l})$ for the original system of noises, and given that we have specified $\xi_{j}=\tilde{\xi}_{j}^{(\nu\mu)}(k)$, the distribution we need to sample from is given by the conditional distribution $\mathbb{G}(\xi_{i}=\tilde{\xi}_{i},\xi_{l}=\tilde{\xi}_{l}|\xi_{j}=\tilde{\xi}_{j}^{(\nu\mu)}$. Using the Schur complement formula, the mean and the covariance matrix of this distribution are given by $\Sigma_{0}\Sigma^{-1}\tilde{\xi}_{j}^{(\nu\mu)}$ and $2\widetilde{D}-\Sigma_{0}\Sigma^{-1}\Sigma_{0}^{T}$ respectively, where
\begin{align}
\Sigma_{0}&= 2\begin{pmatrix}D_{ij}\\D_{lj}\end{pmatrix},\\
\Sigma&= 2D_{jj},\mathrm{~and}\\
\widetilde{D}&= \begin{pmatrix}D_{ii}&&D_{il}\\ D_{il}&&D_{ll}\end{pmatrix}.
\end{align} 
Summarizing, the procedure for generating $X_{i}^{(\mu)}$ and $X_{l}^{(\mu)}$ trajectories for a given $X_{j}^{(\nu)}$ trajectory consists of the following steps at every timestep:
\begin{itemize}
\item The drifts for propagating $X_{i}^{(\mu)}(k)$ and $X_{l}^{(\mu)}(k)$ are obtained using $X_{i}^{(\mu)}(k)$, $X_{j}^{(\nu)}(k)$ and $X_{l}^{(\mu)}(k)$.
\item From the difference between $X_{j}^{(\nu)}(k)$ and $X_{j}^{(\nu)}(k+1)$, $\dot{X}_{j}^{(\nu)}(k)$ is obtained.
\item From $X_{i}^{(\mu)}(k)$ and $X_{l}^{(\mu)}(k)$, together with $X_{j}^{(\nu)}(k)$, the drift in $\dot{X}_{j}^{(\nu)}(k)$ is obtained.
\item From $\dot{X}_{j}^{(\nu)}(k)$ and the drift in $X_{j}$, the noise $\tilde{\xi}_{j}^{(\nu\mu)}(k)$ is obtained.
\item With this noise $\tilde{\xi}_{j}^{(\nu\mu)}(k)$ specified, we can sample the noise $\tilde{\xi}_{i}(k)$ and $\tilde{\xi}_{l}(k)$.
\item Using the drifts and the noise, we propagate $X_{i}^{(\mu)}(k)$ and $X_{l}^{(\mu)}(k)$.
\end{itemize}
Although $X_{i}^{(\mu)}$ and $X_{l}^{(\mu)}$ trajectories become overall a complicated nonlinear function of the $X_{j}^{(\nu)}$ trajectory due to the conditional noise sampling, the computations in each step are linear and simple. For accessing conditional distributions of trajectories of other variables, the reference dynamics is worked out similarly. Thus for every marginal probability computation for the $\nu$-th trajectory, a unique reference dynamics is used, which is fine-tuned to that trajectory. This method of choosing a reference dynamics is a numerical analogue of constructing an approximation for the solution to the stochastic filtering equation\cite{smmoor2023dynamic}, albeit one whose error can be exactly corrected through trajectory reweighting (see also \citen{smgehri2024mutual} for an exact solution to the filtering problem in a class of Poisson-type channels). This drastically reduces the computational cost and makes TE-PWS feasible and accurate.

\section{SM-E: Computing transfer entropy for jump processes}
Here we describe in full detail the computation of transfer entropy in jump processes. For clarity, we reproduce Eqs. 12-14 from main text here, 
\begin{align}
\pi_{X_{i}\to X_{j}}&=-\int_{k\delta t}^{(k+1)\delta t}dt^{'}\lambda_{ij}(t^{'})+\sum_{\alpha=1}^{N_{j}}\ln\mathcal{Q}_{ij}(\alpha)\label{eq:sm_marginal_propensity_1}\\
\pi_{X_{j}}&=-\int_{k\delta t}^{(k+1)\delta t}dt^{'}\lambda_{j}(t^{'})+\sum_{\alpha=1}^{N_{j}}\ln\mathcal{Q}_{j}(\alpha)\label{eq:sm_marginal_propensity_2}\\
\mathcal{T}^{[k]}_{X_{i}\to X_{j}}&=\left\langle \pi_{X_{i}\to X_{j}}-\pi_{X_{j}}\right\rangle \label{eq:sm_jumpte}
\end{align}
where $\pi_{X_{i}\to X_{j}}$ is a functional of the trajectory segments $X_{i,[k,k+1]}$ and $X_{j,[k,k+1]}$, $\pi_{X_{j}}$ is a functional of only the segment $X_{j,[k,k+1]}$, and $\mathcal{T}^{[k]}_{X_{i}\to X_{j}}$ is the transfer entropy from $X_{i}$ to $X_{j}$ over the duration from $k\delta t$ to $(k+1)\delta t$. Here $\alpha$ counts the jumps that change the state of $X_{j}$, $N_{j}$ in number, $\mathcal{Q}_{ij},\lambda_{ij}$ and $\mathcal{Q}_{j},\lambda_{j}$ are jump and escape propensities for $X_{j}$ in the marginal spaces of $(X_{i},X_{j})$ and $(X_{j})$ respectively. The marginal jump propensities $\mathcal{Q}_{ij}(\alpha)$ and $\mathcal{Q}_{j}(\alpha)$ in Eqs. \ref{eq:sm_marginal_propensity_1} and \ref{eq:sm_marginal_propensity_2} are abbreviations for the marginal jump propensities $\mathcal{Q}_{ij}(\alpha,t_{\alpha})$ and $\mathcal{Q}_{j}(\alpha,t_{\alpha})$ with $t_{\alpha}$ denoting the time of the $\alpha$-th jump along the $X_j$ trajectory. The marginal jump propensities for an {\em arbitrary} jump labeled as $\beta$ at an arbitrary time $t$ are defined as
\begin{align}
\mathcal{Q}_{ij}(\beta,t)&\equiv\int D[X_{l,[0,t]}]\;Q_{\beta}(t)\;P(X_{l,[0,t]}|X_{i,[0,t]},X_{j,[0,t]})\label{eq:sm_marginaljumpij}\\
\mathcal{Q}_{j}(\beta,t)&\equiv\int\int D[X_{i,[0,t]}]\;D[X_{l,[0,t]}] \;Q_{\beta}(t)\;P(X_{i,[0,t]},X_{l,[0,t]}|X_{j,[0,t]})\label{eq:sm_marginaljumpj}
\end{align}
where $\mathcal{Q}_{\beta}(t)$ is the jump propensity in the full $d$-dimensional space, and, with a slight abuse of notation, $X_{i,[0,t]}$ denotes the trajectory of $X_{i}$ from time 0 to $t$. The marginal escape propensities $\lambda_{ij}(t)$ and $\lambda_{j}(t)$ in Eqs. \ref{eq:sm_marginal_propensity_1} and \ref{eq:sm_marginal_propensity_2} are defined by summing marginal jump propensities $\mathcal{Q}_{ij}(\beta,t)$ and $\mathcal{Q}_{j}(\beta,t)$ respectively over all possible jumps $\beta$ that change the state of $X_{j}$,
\begin{align}
{\lambda}_{ij}(t)&\equiv\sum_{\beta}{\mathcal{Q}}_{ij}(\beta,t) \\
{\lambda}_{j}(t)&\equiv\sum_{\beta}{\mathcal{Q}}_{j}(\beta,t)
\end{align}

Though the transfer entropy $\mathcal{T}^{[k]}_{X_{i}\to X_{j}}$ is formally defined using both escape and jump propensity terms in Eq. \ref{eq:sm_jumpte}, only the latter terms contribute to the transfer entropy on average, as the escape propensity terms cancel on average, $\langle\lambda_{ij}\rangle=\langle\lambda_{j}\rangle$\cite{smspinney2017transfer}. This is because for any $\beta$-th jump at any time $t$, $\langle\mathcal{Q}_{ij}(\beta,t)\rangle=\langle\mathcal{Q}_{j}(\beta,t)\rangle$ where the angular brackets denote an average over all trajectories, as can be seen from Eqs. \ref{eq:sm_marginaljumpij} and \ref{eq:sm_marginaljumpj}. Thus only the marginal jump propensities at the specific jump times of $X_{j}$, $\mathcal{Q}_{ij}(\alpha)$ and $\mathcal{Q}_{j}(\alpha)$, are formally needed for computing the transfer entropy\cite{smspinney2017transfer}. For computing $\mathcal{Q}_{ij}(\alpha)$ and $\mathcal{Q}_{j}(\alpha)$ by marginalization of the jump propensity in the full $d$-dimensional space, $\mathcal{Q}_{\alpha}(t_{\alpha})$, over conditional distributions, as defined in Eqs. \ref{eq:sm_marginaljumpij} and \ref{eq:sm_marginaljumpj}, we need the conditional probabilities of the trajectories up to the time of each jump $t_{\alpha}$. This is indeed available on-the-fly from the trajectory weights in the RR scheme, as explained in the End Matter (EM) of the main text. The only difference from the case of the diffusive processes is that in the current case, trajectory weights are computed on-the-fly not only up to times $k\delta t$ and $(k+1)\delta t$, but also up to all intermediate times $t_{\alpha}$ that represent the jump times in $X_{j}$. These weights reflect the conditional distributions $P(X_{l,[0,t_{\alpha}]}|X_{i,[0,t_{\alpha}]},X_{j,[0,t_{\alpha}]})$ and $P(X_{i,[0,t_{\alpha}]},X_{l,[0,t_{\alpha}]}|X_{j,[0,t_{\alpha}]})$. We then compute $\mathcal{Q}_{ij}$ and $\mathcal{Q}_{j}$ in Eqs. \ref{eq:sm_marginal_propensity_1} and \ref{eq:sm_marginal_propensity_2} respectively for the $\alpha$-th jump as weighted averages over the trajectories,
\begin{align}
\mathcal{Q}_{ij}(\alpha)&=\int D[X_{l,[0,t_{\alpha}]}]\;Q_{\alpha}(t_{\alpha})\;P(X_{l,[0,t_{\alpha}]}|X_{i,[0,t_{\alpha}]},X_{j,[0,t_{\alpha}]})\\
\mathcal{Q}_{j}(\alpha)&=\int\int D[X_{i,[0,t_{\alpha}]}]\;D[X_{l,[0,t_{\alpha}]}] \;Q_{\alpha}(t_{\alpha})\;P(X_{i,[0,t_{\alpha}]},X_{l,[0,t_{\alpha}]}|X_{j,[0,t_{\alpha}]})
\end{align}
where $Q_{\alpha}(t_{\alpha})$, the jump propensity in the full $d$-dimensional space, is analytically available. Thus, using only the jump propensity terms in Eqs. \ref{eq:sm_marginal_propensity_1}-\ref{eq:sm_jumpte}, we obtain an exact estimate of the transfer entropy, which we call $\mathcal{T}_{X_{i}\to X_{j}}^{(\mathrm{J})}$. This approach of the computation of the transfer entropy using only the jump propensities has been recently shown to have a faster convergence over an alternate time-discretized approach for neural spike train data\cite{smshorten2021estimating}.

\section{SM-F: Reduced variance estimator in jump processes}

Surprisingly, we find that the statistical error in the estimate $\mathcal{T}_{X_{i}\to X_{j}}^{(\mathrm{J})}$ can be further reduced by an order of magnitude by adding to it the integral of the escape terms in Eqs. \ref{eq:sm_marginal_propensity_1} and \ref{eq:sm_marginal_propensity_2}, marginalized with a quadrature of $\delta t$. This reduction arises because fluctuations in the jump terms in Eqs. \ref{eq:sm_marginal_propensity_1} and \ref{eq:sm_marginal_propensity_2} are pathwise anti-correlated to those in the escape terms, even though the latter cancel on average, as explained below.

The improved estimate is obtained by marginalizing the escape terms in Eq. \ref{eq:sm_marginal_propensity_1} and \ref{eq:sm_marginal_propensity_2} over conditional distributions $P\left( X_{l,[0,k]}\big| X_{i,[0,k]},X_{j,[0,k]}\right)$ and $P\left( X_{i,[0,k]},X_{l,[0,k]}\big| X_{j,[0,k]}\right)$ respectively, rather than $P\left( X_{l,[0,t]}\big| X_{i,[0,t]},X_{j,[0,t]}\right)$ and $P\left( X_{i,[0,t]},X_{l,[0,t]}\big| X_{j,[0,t]}\right)$ at every instant of time. Such a quadrature for the marginalization is necessary as the integral in the first terms in Eqs. \ref{eq:sm_marginal_propensity_1} and \ref{eq:sm_marginal_propensity_2} cannot be evaluated in closed form. The improved estimate, based on both the jump propensities J and the escape propensities E, is thus computed as 
\begin{align}
\mathcal{T}^{(\mathrm{J}+\mathrm{E})}_{X_{i}\to X_{j}}\equiv\mathcal{T}^{(\mathrm{J})}_{X_{i}\to X_{j}}&-\int_{k\delta t}^{(k+1)\delta t}dt^{'}{\lambda}_{ij}(t^{'})+\int_{k\delta t}^{(k+1)\delta t}dt^{'}{\lambda}_{j}(t^{'})\label{eq:T_JpE}
\end{align}
where $\mathcal{T}^{(\mathrm{J})}_{X_{i}\to X_{j}}$ is the estimate from the previous section based on only the jump propensity terms in Eqs. \ref{eq:sm_marginal_propensity_1} and \ref{eq:sm_marginal_propensity_2}. In Eq. \ref{eq:T_JpE}, ${\lambda}_{ij}(t)$ and ${\lambda}_{j}(t)$ are computed as ${\lambda}_{ij}(t)\equiv\sum_{\beta}{\mathcal{Q}}_{ij}(\beta,t)$, ${\lambda}_{j}(t)\equiv\sum_{\beta}{\mathcal{Q}}_{j}(\beta,t)$ with $\mathcal{Q}_{ij}(\beta,t)$ and ${\mathcal{Q}}_{j}(\beta,t)$ computed from Eqs. \ref{eq:sm_marginaljumpij} and \ref{eq:sm_marginaljumpj}, respectively, but with 
\begin{align}
{P}(X_{l,[0,t]}|X_{i,[0,t]},X_{j,[0,t]})&\simeq P(X_{l,[0,k]}|X_{i,[0,k]},X_{j,[0,k]})\;P_{0}(X_{l,[k,t]}|X_{l,[0,k]}) \label{eq:sm_approxcondij}\\
{P}(X_{i,[0,t]},X_{l,[0,t]}|X_{j,[0,t]})&\simeq P(X_{i,[0,k]},X_{l,[0,k]}|X_{j,[0,k]})\;P_{0}(X_{i,[k,t]},X_{l,[k,t]}|X_{i,[0,k]},X_{l,[0,k]}) \label{eq:sm_approxcondj}
\end{align}
Here $X_{i,[0,k]}$ and $X_{i,[k,t]}$ denote the $X_{i}$ trajectory between times 0 and $k\delta t$ and between times $k\delta t$ and $t$ respectively; $P_0$ is, as before, the reference distribution from which we generate the trajectories of the hidden variables. Hence, we see that the estimator $\mathcal{T}^{(\mathrm{J}+\mathrm{E})}_{X_{i}\to X_{j}}$ is computed from the trajectory weights at time $k\delta t$ as available from the RR scheme, rather than the weight at every instant of time $t$. We refer to this as the quadrature approximation for marginalization, which incurs an $\mathcal{O}(\delta t)$ error.

This estimate $\mathcal{T}_{X_{i}\to X_{j}}^{(\mathrm{J}+\mathrm{E})}$ has a significantly smaller variance than the estimate $\mathcal{T}_{X_{i}\to X_{j}}^{(\mathrm{J})}$. This is because in the former, fluctuations in the logarithm of jump propensities are suppressed by anti-correlated fluctuations in escape propensities. Physically, if a jump fires more than average, the waiting times between the jumps become that much more improbable. We can show this by calculating the fluctuations in $\sum_{\alpha}\ln [Q_{ij}(\alpha)/Q_{j}(\alpha)]-\int{dt^{'}} (\lambda_{ij}-\lambda_{j})$ as below. Consider the firing of only one kind of jump of $X_{j}$ with an average propensity $\mathcal{Q}_{j}=\mathcal{Q}^{*}$, \textit{i.e.}, an average escape propensity of $\lambda_{j}=\mathcal{Q}^{*}$. When we condition on a specific $X_{i}$ trajectory, the fluctuation of the latter affects the conditional jump propensity for $X_{j}$, \textit{i.e.}, $\mathcal{Q}_{ij}$ deviates from its mean $\mathcal{Q}^{*}$, resulting in information transfer from $X_{i}$ to $X_{j}$. An additional source of fluctuations in the trajectory of $X_{j}$ is the stochastic number of times the jump fires within $\delta t$, $N_{j}$. Over a small $\delta t$, when $Q_{ij}$ stays temporally almost constant, $N_{j}$ is Poisson-distributed with a mean $Q_{ij}\delta t$. The transfer entropy between times $k\delta t$ and $(k+1)\delta t$ using Eq. \ref{eq:sm_jumpte} then is
\begin{align}\label{eq:tejumpwait}
\mathcal{T}^{[k]}_{X_{i}\to X_{j}}&=\Bigg\langle Q_{ij}\delta t\ln\frac{Q^{ij}}{Q^{*}}-(Q_{ij}-Q^{*})\delta t \Bigg\rangle
\end{align}
where the angular brackets now denote an average over $X_{i}$ trajectories.
We see that the second term fluctuates around a mean of 0. By Taylor expanding the first term upto second order in $(Q_{ij}-Q^{*})$, we find
\begin{align}
\mathcal{T}^{[k]}_{X_{i}\to X_{j}}&= \Bigg\langle (Q_{ij}-Q^{*})\delta t+\frac{\delta t(Q_{ij}-Q^{*})^{2}}{2Q^{*}}+\mathcal{O}\big( (Q_{ij}-Q^{*})^{3}\big) -(Q_{ij}-Q^{*})\delta t \Bigg\rangle \\
&=\Bigg\langle \frac{\delta t(Q_{ij}-Q^{*})^{2}}{2Q^{*}}+\mathcal{O}\big( (Q_{ij}-Q^{*})^{3}\big) \Bigg\rangle 
\end{align}
So fluctuations in the second term of Eq. \ref{eq:tejumpwait} cancel a part of the fluctuations in the first term. This results in a smaller overall variance compared to the case where only the first term of Eq. \ref{eq:tejumpwait} is used. The above proof also holds in the case of multiple kinds of jumps in $X_{j}$, such as increments and decrements in copy numbers, because the escape propensities $\lambda_{ij}$ and $\lambda_{j}$ are additive over the different kinds of jumps and the above proof works separately for each kind of jump.

\begin{figure}[H]
\centering
\includegraphics[width=15cm]{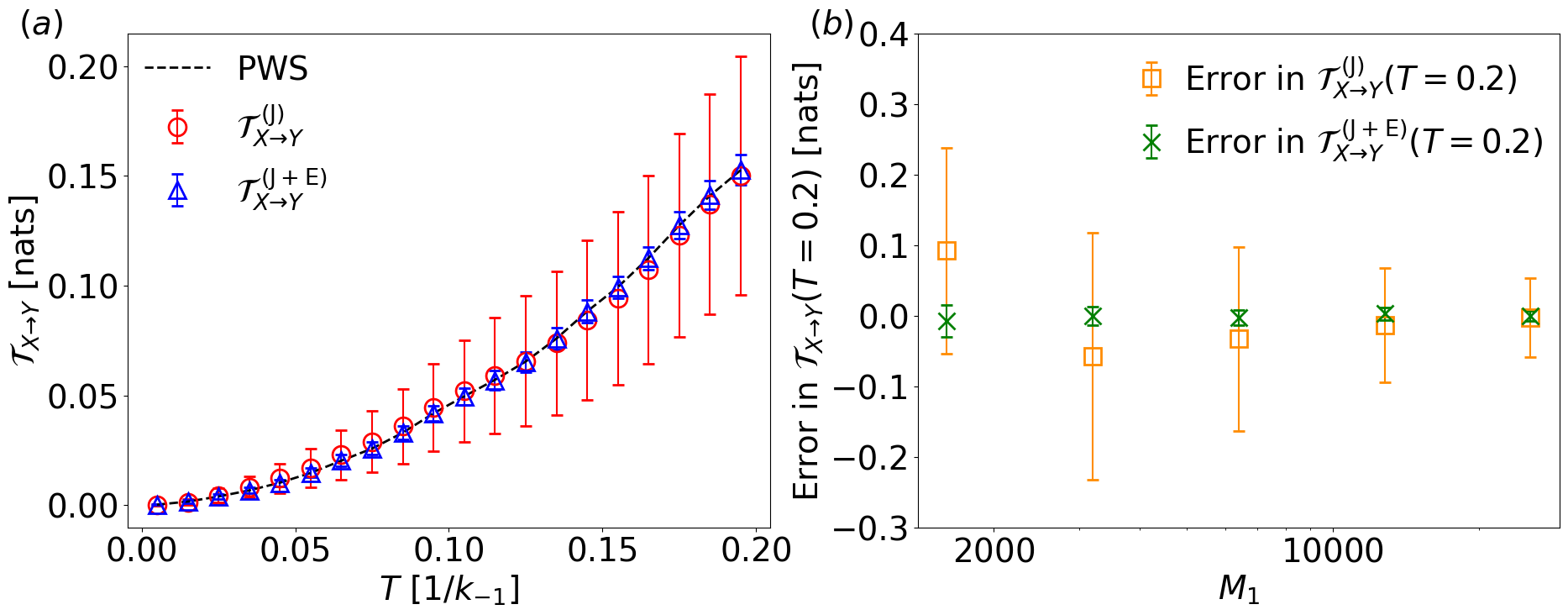}
\caption{Accuracy of transfer entropy estimates $\mathcal{T}_{X_{i}\to X_{j}}^{(\mathrm{J})}$ and $\mathcal{T}_{X_{i}\to X_{j}}^{(\mathrm{J}+\mathrm{E})}$ in jump processes. (a) Transfer entropy as a function of time from the jump-based estimate $\mathcal{T}_{X\to Y}^{(\mathrm{J})}$ (red circles), reduced-variance estimate $\mathcal{T}_{X\to Y}^{(\mathrm{J}+\mathrm{E})}$ (blue triangles) and \textit{ground truth} estimate from PWS (black dashed line) in the chemical reaction network with two species $X$ and $Y$. (b) Error at trajectory duration $T=0.2/k_{-1}$, defined as the difference of each estimate from the PWS estimate, as a function of increasing Monte-Carlo averaging. Plotted are errors in the jump-based estimate $\mathcal{T}_{X\to Y}^{(\mathrm{J})}$ (orange squares) and the reduced-variance estimate $\mathcal{T}_{X\to Y}^{(\mathrm{J}+\mathrm{E})}$ (green crosses). Initial conditions for the simulations are $n_{X}=50$ and $n_{Y}=500$, where $n_{X}$ and $n_{Y}$ are the number of species of $X$ and $Y$ respectively. $M_{1}$ for (a) is 25600. $M_{2}=1000$ for both subfigures.}
\label{fig:jump}
\end{figure}

We numerically demonstrate this effect in a chemical reaction network of two species $X$ and $Y$, consisting of reactions $\phi \rightarrow X, X\rightarrow\phi, X\rightarrow X+Y, Y\rightarrow\phi$, with rate constants $k_{1}=50$, $k_{-1}=1$, $k_{2}=10$ and $k_{-2}=10$ respectively. Plotted in Figs. \ref{fig:jump}a and b are the two different transfer entropy estimates and their errors as a function of increasing trajectory duration and increasing statistical averaging respectively. As there is no feedback, the transfer entropy is formally equal to the exact mutual information estimate from PWS\cite{smreinhardt2023path}, against which we have compared our results. We find that though both estimates yield unbiased results, the reduced-variance estimate $\mathcal{T}_{X\to Y}^{(\mathrm{J}+\mathrm{E})}$ has an order of magnitude smaller error than the jump-based estimate $\mathcal{T}_{X\to Y}^{(\mathrm{J})}$, for the same computational cost. On the other hand, the $\mathcal{O}(\delta t)$ error in $\mathcal{T}_{X\to Y}^{(\mathrm{J}+\mathrm{E})}$ from the quadrature in marginalization is negligible. $\mathcal{T}_{X_{i}\to X_{j}}^{(\mathrm{J}+\mathrm{E})}$ is thus a more accurate estimate of the transfer entropy than $\mathcal{T}_{X_{i}\to X_{j}}^{(\mathrm{J})}$. We expect this theoretical result to be tested using experimental data in the future, such as using data from neural spike trains\cite{smshorten2021estimating}.

\section{SM-G: Variants of transfer entropy}

A central advance in TE-PWS is the computation of the probability of a dynamical fluctuation by marginalization over trajectories of hidden variables. Here we show, in the context of discretized Langevin processes, how the same approach can be used to compute other trajectory-based metrics of directional information transmission beyond Schreiber's transfer entropy.

\textbf{Directed information.} An alternate measure of information transmission from the trajectory $X_{i,[0,N]}$ to $X_{j,[0,N]}$ is given by directed information\cite{smmassey1990causality}, defined as
\begin{align}\label{eq:directedinfo}
I[X_{i,[0,N]}\to X_{j,[0,N]}]= \sum_{k=0}^{N-1}I\left( X_{j}(k+1);X_{i,[0,k+1]}|X_{j,[0,k]}\right) &=\Bigg\langle \ln\frac{\prod_{k}P\left( X_{j}(k+1)|X_{i,[0,k+1]},X_{j,[0,k]}\right) }{\prod_{k} P\left( X_{j}(k+1)|X_{j,[0,k]}\right) } \Bigg\rangle 
\end{align}
By comparing the above definition with that of the transfer entropy in Eq. 2 of the main text, we see that the directed information incorporates the stepwise information transmission into $X_{j}$ coming from the entire trajectory of $X_{i}$ including the current value $X_{i}(k+1)$, while Schreiber's transfer entropy excludes the flow from the current value of $X_{i}$ (compare Eq. \ref{eq:directedinfo} against Eqs. 2 and 5 in the main text). We can bring the directed information to a computable form by taking $X_{i}(k+1)$ out of the conditioning in the numerator, 
\begin{align}
I[X_{i,[0,N]}\to X_{j,[0,N]}]&= \Bigg\langle \ln\frac{\prod_{k}P\left( X_{j}(k+1)|X_{i,[0,k+1]},X_{j,[0,k]}\right) }{\prod_{k} P\left( X_{j}(k+1)|X_{j,[0,k]}\right) } \Bigg\rangle \nonumber\\
&=\Bigg\langle \ln\prod_{k}P\left( X_{i}(k+1),X_{j}(k+1)|X_{i,[0,k]},X_{j,[0,k]}\right) \Bigg\rangle -\Bigg\langle \ln\prod_{k} P\left( X_{j}(k+1)|X_{j,[0,k]}\right) \Bigg\rangle \nonumber\\
&\qquad\qquad-\Bigg\langle \ln\prod_{k} P\left( X_{i}(k+1)|X_{i,[0,k]},X_{j,[0,k]}\right)  \Bigg\rangle \label{eq:directedinfo_decomp}
\end{align}
Here we note that only two marginalization integrals are actually required for computing the three probabilities in Eq. \ref{eq:directedinfo_decomp}. One is over the conditional distribution $P(X_{l,[0,k]}|X_{i,[0,k]},X_{j,[0,k]})$ and the other over $P(X_{i,[0,k]},X_{l,[0,k]}|X_{j,[0,k]})$, exactly the same as those sampled in TE-PWS for calculating $\mathcal{T}_{X_{i}\to X_{j}}$, where $X_{l}$ denote all variables except $X_{i}$ and $X_{j}$. These two distributions can give the probabilities in Eq. \ref{eq:directedinfo_decomp} as
\begin{align}
P\left( X_{i}(k+1),X_{j}(k+1)|X_{i,[0,k]},X_{j,[0,k]}\right) &=\int D[X_{l,[0,k]}]\;P\left( X_{l,[0,k]}|X_{i,[0,k]},X_{j,[0,k]}\right) \nonumber\\
&\qquad.\;P\left( X_{i}(k+1),X_{j}(k+1)|X_{i,[0,k]},X_{j,[0,k]},X_{l,[0,k]}\right) \label{eq:directedinfo_mc1}\\
P\left( X_{i}(k+1)|X_{i,[0,k]},X_{j,[0,k]}\right) &=\int D[X_{l,[0,k]}]\;P\left( X_{l,[0,k]}|X_{i,[0,k]},X_{j,[0,k]}\right) \nonumber\\
&\qquad.\;P\left( X_{i}(k+1)|X_{i,[0,k]},X_{j,[0,k]},X_{l,[0,k]}\right) \label{eq:directedinfo_mc2}\\
P\left( X_{j}(k+1)|X_{j,[0,k]}\right) &=\int\int D[X_{i,[0,k]}]\;D[X_{l,[0,k]}]\;P\left( X_{i,[0,k]},X_{l,[0,k]}|X_{j,[0,k]}\right) \nonumber\\
&\qquad.\;P\left( X_{j}(k+1)|X_{i,[0,k]},X_{j,[0,k]},X_{l,[0,k]}\right) \label{eq:directedinfo_mc3}
\end{align}
where, aside from the two conditional distributions $P(X_{l,[0,k]}|X_{i,[0,k]},X_{j,[0,k]})$ and $P(X_{i,[0,k]},X_{l,[0,k]}|X_{j,[0,k]})$, all the other probabilities in the integrals are analytically available.
Thus, by sampling the two conditional distributions through the RR scheme, computing the three averages in Eqs. \ref{eq:directedinfo_mc1}-\ref{eq:directedinfo_mc3} as averages over those conditional distributions, plugging them in Eq. \ref{eq:directedinfo_decomp}, and evaluating Eq. \ref{eq:directedinfo_decomp} as a single Monte-Carlo average, TE-PWS can compute directed information with the same computational cost as transfer entropy.

\textbf{Conditional transfer entropy.} The conventional transfer entropy can have a positive value even when there is no direct causal link from the input to the output variable, when information is being causally transmitted through intermediate variables. This motivated the definition of a conditional transfer entropy, also known as causation entropy, that can measure direct causal links\cite{smsun2014causation,smjames2016information,smfaes2016information}. For any choice of a third variable $X_{m}$, the conditional transfer entropy is defined as
\begin{align}
&\mathcal{T}_{X_{i}\to X_{j}|X_{m}}=\sum_{k=0}^{N-1}I\left( X_{j}(k+1);X_{i,[0,k]}|X_{j,[0,k]},X_{m,[0,k]}\right) =\sum_{k}\Bigg\langle \ln\frac{P\left( X_{j}(k+1)|X_{i,[0,k]},X_{j,[0,k]},X_{m,[0,k]}\right) }{ P\left( X_{j}(k+1)|X_{j,[0,k]},X_{m,[0,k]}\right) } \Bigg\rangle \label{eq:conditionedtransfer}
\end{align}
This expectation can be computed similar to the ordinary transfer entropy in Eq. 5 of the main text. The average is computed in a Monte-Carlo fashion over simulated trajectories of all variables. For each set of trajectories, the numerator and denominator are computed by marginalizing over all other hidden variables $X_{l}$ which exclude $X_{m}$ this time. The optimal reference dynamics should now be chosen to incorporate the effects of the $X_{m}$ trajectory through a frozen field of drift and diffusion resulting from the fixed $X_{m}$ trajectory, similar to how the $X_{i}$ and $X_{j}$ trajectories influence the reference dynamics as discussed above, under {\bf Choice of reference distribution}.
Thus, calculation of each conditional transfer entropy with TE-PWS requires two marginalization integrals, similar to the ordinary transfer entropy.

\textbf{Filtered transfer entropy.} Recently, filtered transfer entropy has been proposed as a way to quantify information transfer in the spirit of filtering theory\cite{smchetrite2019information}. The filtered transfer entropy from $X_{i}$ to $X_{j}$ is defined as
\begin{align}
&\widehat{\mathcal{T}}_{X_{i}\to X_{j}}=\sum_{k=0}^{N-1}I\left( X_{i}(k+1);X_{j}(k+1)|X_{j,[0,k]}\right) \nonumber\\
&=\sum_{k}\Bigg\langle \ln\frac{P\left( X_{i}(k+1),X_{j}(k+1)|X_{j,[0,k]}\right) }{P\left( X_{i}(k+1)|X_{j,[0,k]}\right) P\left( X_{j}(k+1)|X_{j,[0,k]}\right) } \Bigg\rangle \label{eq:filtered_transfer}
\end{align}
which quantifies how much the prediction of $X_{i}(k+1)$ is improved by using $X_{j}(k+1)$ in addition to the past trajectory $X_{j,[0,k]}$. The computation of $\widehat{\mathcal{T}}_{X_{i}\to X_{j}}$ requires marginalization over only one conditional distribution, $P(X_{i,[0,k]},X_{l,[0,k]}|X_{j,[0,k]})$. Each of the probabilities in Eq. \ref{eq:filtered_transfer} can be computed by averaging analytically available transition probabilities $P\left( X_{i}(k+1),X_{j}(k+1)|X_{i,[0,k]},X_{j,[0,k]},X_{l,[0,k]}\right)$, $P\left( X_{i}(k+1)|X_{i,[0,k]},X_{j,[0,k]},X_{l,[0,k]}\right)$ and $P\left( X_{j}(k+1)|X_{i,[0,k]},X_{j,[0,k]},X_{l,[0,k]}\right)$, over this conditional distribution $P(X_{i,[0,k]},X_{l,[0,k]}|X_{j,[0,k]})$, which is provided by TE-PWS through the RR scheme. Thus, TE-PWS can be used to compute filtered transfer entropy at half the computational cost as Schreiber's transfer entropy.

\section{SM-H: Comparison with approximate methods}
Currently many approximate methods are being used to estimate the transfer entropy rate. The accuracy of these approximations have not been tested so far in the absence of an exact technique. Here we perform an extensive comparison of the performance of these techniques with exact estimates from TE-PWS. We thereby demonstrate for the first time that these methods exhibit significant systematic errors in both linear and nonlinear systems, and that no approximate technique is accurate across all systems. Combined with the demonstration in the main text that the computational cost of TE-PWS is either comparable or significantly cheaper compared with the approximate techniques, our conclusion is that TE-PWS should be the method of choice whenever a dynamical model for the system of interest is known.

First we review the most widely used approximate methods for computing transfer entropies. All existing methods truncate the history dependence of the transfer entropy to $(k+1)$ time points in the past. Hence, we introduce, following Schreiber\cite{smschreiber2000measuring}, the $k$-truncated transfer entropy rate from $X_{i}$ to $X_{j}$ in steady-state,
\begin{align}
\dot{\mathcal{T}}_{X_{i}\to X_{j}}^{[k]}&=\frac{1}{\delta t}I\left( X_{j}(k+1);X_{i,[0,k]}|X_{j,[0,k]}\right)
\end{align}
which corresponds to the $k$-th term in the sum in Eq. 3 of the main text, divided by $\delta t$.
Here $(k+1)\geq1$ is the number of past snapshots of $X_{i}$ and $X_{j}$ replacing the full steady-state trajectory $X_{i,[0,\infty]}$ and $X_{j,[0,\infty]}$ (see Fig. 1 of the main text); indeed, for systems in steady-state, the $k$-dependent terms in Eq. 3 of the main text all converge to the same value for history lengths $(k+1)\delta t$ longer than the largest relaxation time in the system, yielding the transfer entropy rate $\dot{\mathcal{T}}_{X_{i}\to X_{j}}$ when divided by $\delta t$.
We note that $k=0$ produces the one-step truncated transfer entropy approximation,
\begin{align}
\dot{\mathcal{T}}_{X_{i}\to X_{j}}\approx\dot{\mathcal{T}}_{X_{i}\to X_{j}}^{[k=0]}&=\frac{1}{\delta t}I\left( X_{j}(1);X_{i}(0)|X_{j}(0)\right) \label{eq:onestepTE}\\
&=\frac{1}{\delta t}\int dX_{j}(1)\;dX_{i}(0)\;dX_{j}(0)\;P(X_{j}(1),X_{i}(0),X_{j}(0))\ln\frac{P(X_{j}(1)|X_{i}(0),X_{j}(0)}{P(X_{j}(1)|X_{j}(0))} \label{eq:onestepTE_eq}
\end{align}
where the history dependence is ignored. This approximation is widely used due to its relatively small computational cost compared to that of obtaining an estimate with a higher value of $k$\cite{smschreiber2000measuring,smpahle2008information,smlahiri2017information,smchetrite2019information}.
Nevertheless, it is well-recognized in the literature that the truncation of the trajectory history-length to a value shorter than the longest correlation time of the system does not accurately capture the information transfer\cite{smvicente2011transfer}. Indeed, the correct procedure is to evaluate $\dot{\mathcal{T}}_{X_{i}\to X_{j}}^{[k]}$ for increasing values of $k$ till the estimate becomes independent of $k$, \textit{i.e.}, converges to $\dot{\mathcal{T}}_{X_{i}\to X_{j}}$, 
\begin{align}
\dot{\mathcal{T}}_{X_{i}\to X_{j}}=\lim_{k\to\infty}\dot{\mathcal{T}}_{X_{i}\to X_{j}}^{[k]}\label{eq:TElimit}
\end{align}

The brute-force approach, which is based on computing the transfer entropy by estimating the necessary probability distribution $P(X_{j}(k+1),X_{i,[0,k]},X_{j,[0,k]})$ via histogram binning, becomes intractable for large $k$. In particular, for obtaining a converged estimate, the required amount of data explodes exponentially with $k$ as $N_{x}^{2(k+1)+1}$, where $N_{x}$ is the number of bins used for each time point of $X_{i}$ or $X_{j}$. Thus it becomes computationally prohibitive to estimate the transfer entropy by simply binning signal trajectories.

\textbf{Gaussian framework.} Hence, further approximations are typically being made. These approximate techniques assume simple forms for the probability distribution of the signal trajectories. One popular approximation is to assume that the probability distribution is Gaussian\cite{smtostevin2009mutual,smbarnett2014mvgc}. Then the approximate transfer entropy rate $\dot{\mathcal{T}}_{X_{i}\to X_{j}}^{[k]}$ has a simple form,
\begin{align}\label{eq:GaussianTE}
\dot{\mathcal{T}}_{X_{i}\to X_{j}}=\lim_{k\to\infty}\dot{\mathcal{T}}_{X_{i}\to X_{j}}^{[k]}\approx \lim_{k\to\infty}\frac{1}{2\delta t}\ln\frac{\sigma^{2}_{X_{j}(k+1)|X_{j,[0,k]}}}{\sigma^{2}_{X_{j}(k+1)|X_{i,[0,k]},X_{j,[0,k]}}}\equiv\lim_{k\to\infty}\frac{1}{2\delta t}\ln\frac{\sigma^{2}_{X_{j+}|X_{j-}}}{\sigma^{2}_{X_{j+}|X_{i-},X_{j-}}}
\end{align}
where the numerator inside the logarithm is the conditional variance of $X_{j}(k+1)$ when only the trajectory $X_{j,[0,k]}$ is known, the denominator is the conditional variance of $X_{j}(k+1)$ when both $X_{i,[0,k]}$ and $X_{j,[0,k]}$ are known, and $X_{j+}$, $X_{i-}$ and $X_{j-}$ are abbreviations for $X_{j}(k+1)$, $X_{i,[0,k]}$ and $X_{j,[0,k]}$ respectively. The approximation becomes exact in the case of a linear system, \textit{i.e.}, if the stochastic dynamics is an OU process, but introduces an uncontrolled error in the presence of nonlinearity, such as in model D where the drift terms are nonlinear. In practice, the approximation is implemented by empirically estimating the $[2(k+1)+1]\times[2(k+1)+1]$-dimensional variance-covariance matrix $\Sigma_{\mathbb{G}}$ of the joint probability distribution $P(X_{j+},X_{i-},X_{j-})$ from steady-state signal trajectories. The conditional variances in Eq. \ref{eq:GaussianTE} are then computed using the Schur complement formula\cite{smbarnett2009granger,smhahs2013transfer},
\begin{align}
\sigma^{2}_{X_{j+}|X_{j-}}&=\sigma^{2}_{X_{j+}}-\mathbf{\Sigma}_{X_{j+};X_{j-}}\mathbf{\Sigma}_{X_{j-};X_{j-}}^{-1}\mathbf{\Sigma}_{X_{j+};X_{j-}}^{T}\label{eq:schur1}\\
\sigma^{2}_{X_{j+}|X_{i-},X_{j-}}&=\sigma^{2}_{X_{j+}}-\mathbf{\Sigma}_{X_{j+};(X_{i-},X_{j-})}\mathbf{\Sigma}_{(X_{i-},X_{j-});(X_{i-},X_{j-})}^{-1}\mathbf{\Sigma}_{X_{j+};(X_{i-},X_{j-})}^{T}\label{eq:schur2}
\end{align}
where $\mathbf{\Sigma}_{X_{j+};X_{j-}}$ is the $1\times (k+1)$-dimensional covariance vector of $X_{j}(k+1)$ with the trajectory $X_{j,[0,k]}$, $\mathbf{\Sigma}_{X_{j+};(X_{i-},X_{j-})}$ is the $1\times2(k+1)$-dimensional covariance vector of $X_{j}(k+1)$ with the trajectory pair $(X_{i,[0,k]},X_{j,[0,k]})$, $\mathbf{\Sigma}_{X_{j-};X_{j-}}$ is the $(k+1)\times (k+1)$-dimensional variance-covariance matrix of the trajectory $X_{j,[0,k]}$ with itself, and  $\mathbf{\Sigma}_{(X_{i-},X_{j-});(X_{i-},X_{j-})}$ is the $2(k+1)\times2(k+1)$ dimensional variance-covariance matrix of the trajectory pair $(X_{i,[0,k]},X_{j,[0,k]})$ with itself. All these covariance matrices and vectors are simply submatrices and columns of the full variance-covariance matrix $\Sigma_{\mathbb{G}}$ of the full probability distribution $P(X_{j+},X_{i-},X_{j-})$. This computation of $\dot{\mathcal{T}}_{X_{i}\to X_{j}}^{[k]}$ needs to be done for increasing values of $k$ such that a converged $k\to\infty$ limit can be obtained. The cost of computing each $\dot{\mathcal{T}}_{X_{i}\to X_{j}}^{[k]}$ from the signal trajectories is then two-fold: the cost of empirically estimating $\Sigma_{\mathbb{G}}$ which scales as $\mathcal{O}(k^{2})$, and the cost of performing the matrix operations in Eqs. \ref{eq:schur1} and \ref{eq:schur2}, especially the matrix inversions, which scale as $\mathcal{O}(k^{3})$.

\begin{figure}[t]
\centering
\includegraphics[width=15cm]{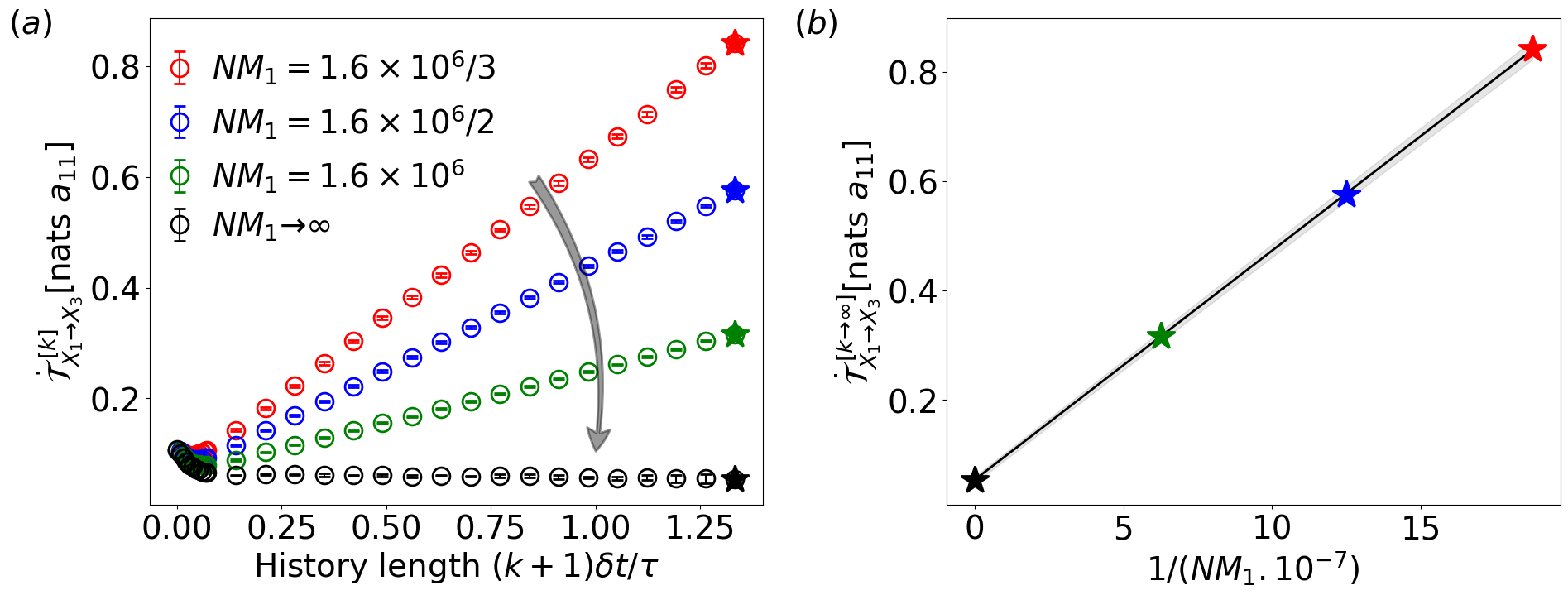}
\caption{Convergence of the Gaussian transfer entropy rate estimate in the nonlinear model D, which has a relaxation timescale of $\tau=60 a_{11}^{-1}$. (a) The transfer entropy rate $\dot{\mathcal{T}}_{X_{1}\to X_{3}}^{[k]}$ as a function of the history length $(k+1)$, for different choices of $NM_{1}$. For each value of $k$, the estimates from different $NM_{1}$ were linearly extrapolated to the infinite data limit ($1/(NM_{1})\to0$) to obtain the black circles. The black arrow is an aid to the eye for the convergence with increasing $NM_{1}$. The linear extrapolation is plotted in (b) for the largest $k$ (stars). The black star is interpreted as the converged transfer entropy rate estimate $\dot{\mathcal{T}}_{X_{1}\to X_{3}}$ in the limit of infinite data.}
\label{fig:gaussian}
\end{figure}
Fig. \ref{fig:gaussian} shows the protocol for obtaining a converged transfer entropy estimate with the Gaussian framework in the infinite data limit for the nonlinear model D. First, keeping the total amount of data $NM_{1}$ fixed, \textit{i.e.}, for a trajectory of total duration $NM_{1}\delta t$, we use the Gaussian framework to estimate the transfer entropy rate for increasing values of history length $(k+1)$. We find that at large $k$, there is a significant data-size dependent bias in the estimate. For every value of $k$, we correct for this bias by linearly extrapolating the estimates from different $NM_{1}$ to the infinite data limit, $1/(NM_{1})\to 0$. This extrapolation gives a transfer entropy rate that is converged with respect to the history length $(k+1)$ and data size $NM_{1}$. This same converged value has been reported in the Table 1 in the main text. The protocol for the results in the linear system (model A), reported in Fig. 2 and Table 1 of the main text, was similar, with the convergence being studied upto a value of $k=3200$.

\begin{figure}[b]
\centering
\includegraphics[width=16cm]{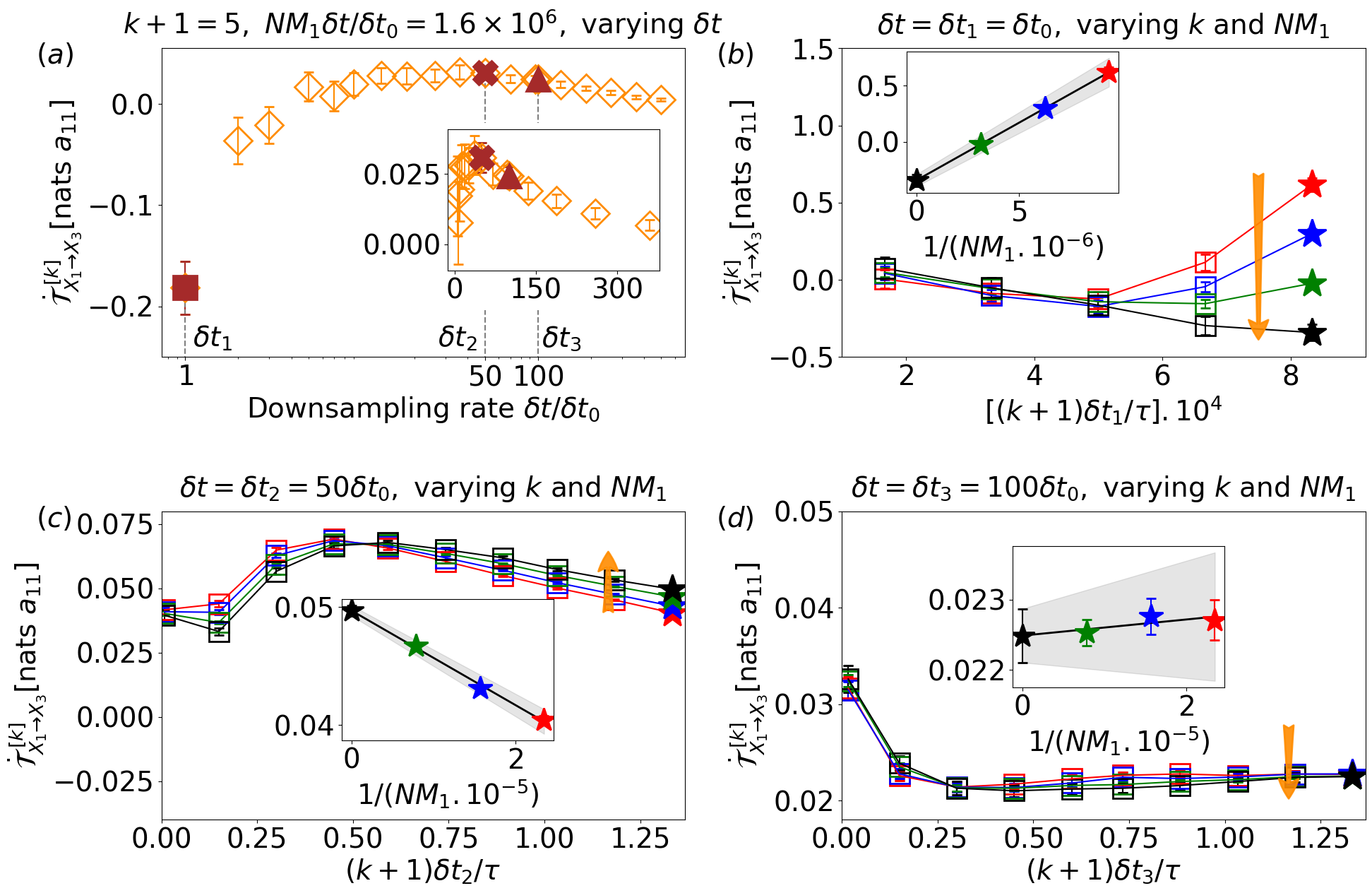}
\caption{Protocol for obtaining a converged transfer entropy estimate from the KSG algorithm in the $k\to\infty$ limit in the nonlinear model D, which has a relaxation timescale of $\tau=60a_{11}^{-1}$. The number of nearest neighbor data points in the KSG algorithm has been kept fixed at $k^{'}=4$. (a) Keeping $k+1=5$ and the total trajectory duration fixed, the simulated trajectories are downsampled with a rate $\delta t/\delta t_{0}$, where $\delta t_{0}=0.01 a_{11}^{-1}$ is the simulation timestep. The transfer entropy rate estimate has an unphysical negative value without downsampling ($\delta t/\delta t_{0}=1$). With increasing downsampling, the estimate increases to a peak and then decays to 0 as shown in the same data plotted with a logarithmic x-scale in (a) and with a linear x-scale in its inset. The decay of the transfer entropy rate with increasingly coarse downsampling is expected due to the decorrelation between temporally distant configurations in the downsampled trajectory. The vertical dashed lines mark the three choices of downsampling rates investigated below. (b) Keeping $\delta t=\delta t_{1}=\delta t_{0}$, \textit{i.e.}, without any downsampling and corresponding to the brown square from panel (a), $k$ is varied for different fixed values of $NM_{1}$, given by $NM_{1}=1.6\times 10^{6}$ (green symbols), and half and one-third of that value (blue and red symbols respectively). For each $k$, the values at different $NM_{1}$ are then extrapolated linearly to the infinite data limit, $1/(NM_{1})\to0$. Lines connecting the symbols and the orange arrow are aids to the eye to show the trend with increasing $k$ and increasing $NM_{1}$ respectively. The linear fit for the highest $k$ is shown in the inset with star symbols. Even the extrapolated values to the infinite data limit are unphysically negative. (c) Similar to (b) but for a downsampling rate of $\delta t=\delta t_{2}=50\delta t_{0}$ corresponding to the brown crosses from panel (a). The transfer entropy rate here does not converge to a number independent of $k$ even when extrapolated to the infinite data limit. (d) Similar to (b) but for a downsampling rate of $\delta t=\delta t_{3}=100\delta t_{0}$ corresponding to the brown triangles from panel (a). The transfer entropy rate converges as a function of $k$ to a physically possible value on extrapolation to the infinite data limit. The converged rate estimate $\dot{\mathcal{T}}_{X_{1}\to X_{3}}$ is then the black star symbol. This rate is reported in Table 1 of the main text.}
\label{fig:ksg}
\end{figure}

\textbf{KSG algorithm.} For nonlinear systems, a Gaussian approximation is usually not expected to be accurate. Then a different approximation, broadly known as a $k$-nearest neighbor ($k$-NN) entropy estimate, is often used to calculate the transfer entropy without having to compute the full histogram of $P(X_{j+},X_{i-},X_{j-})$.\cite{smkraskov2004estimating} By unfortunate convention, the $k$ in $k$-NN would usually refer to the number of nearest neighbors in the hyperspace of all data, not to the history length. For resolving the ambiguity in the notation $k$, we will henceforth call the number of nearest neighbors $k^{'}$ and reserve, following Schreiber,\cite{smschreiber2000measuring} $(k+1)$ for the history length of the trajectory, similar to the previous paragraphs.  Then in the nearest neighbor-based approach, the probability density value $P(X_{j+},X_{i-},X_{j-})$ is locally estimated around each sample data point by computing the distance to a few nearest data points in the $[2(k+1)+1]$-dimensional space. The approximation is that the probability density is uniform within the smallest volume that encompasses the first $k^{'}$ nearest neighbors, where $k^{'}$ is a hyperparameter. In the limit of infinite data, the sample points are infinitely dense, so the volume encompassing $k^{'}$ nearest neighbors becomes infinitesimally small, and at that scale any continuous probability density would appear uniform. Hence, the estimate is asymptotically unbiased in the large data limit. The first such estimate was the Kozachenko-Leonenko (KL) estimate\cite{smkozachenko1987sample,smsingh2003nearest} where the volume encompassing the $k^{'}$-nearest neighbors is a hypersphere. An improved estimate was subsequently formulated by Kraskov, St\"ogbauer and Grassberger (KSG) who improved the biases of the KL-estimator by a cancellation of errors from a clever choice of the shape encompassing the $k^{'}$-nearest neighbors\cite{smkraskov2004estimating}. The KSG estimator is currently widely used to estimate the transfer entropy through Eq. \ref{eq:TElimit} directly from experimental data without any model assumptions\cite{smwibral2011transfer,smwollstadt2014efficient,smwibral2014directed,smlizier2014jidt}. It has been incorporated into several openly available toolkits such as TRENTOOL\cite{smlindner2011trentool} and JIDT\cite{smlizier2014jidt}. A comprehensive and helpful guide for understanding and implementing the KSG estimate for transfer entropy is in \citen{smwibral2014directed}. 

Despite its wide use, it is well-recognized that the KSG estimate suffers from large sample-size dependent biases when the sample size is small due to the assumption that the probability density is locally uniform\cite{smholmes2019estimation}. These biases become prominent when the sampling is non-uniform, for example due to bistability, when the sample data points are not independent, when the information source and sink are strongly correlated, or when the dimension of the trajectories is large due to the history length $(k+1)$ being large\cite{smvicente2011transfer,smgao2015efficient,smgao2018demystifying}. The biases in the estimator can be so large as to give statistically significant negative values for the mutual information, which is physically meaningless\cite{smholmes2019estimation,smmarx2022estimating}. Moreover, the sample-size dependent bias in the KSG estimate can not be easily corrected as the scaling of the bias with sample size is not known\cite{smholmes2019estimation}. Hence to prevent large errors in the KSG estimate, a veritable zoo of hyperparameters has been invented, such as distinct history lengths for the source and the sink trajectories $X_{i-}$ and $X_{j-}$, a tunable delay between the past trajectories $(X_{i-},X_{j-})$ and the future state $X_{i+}$, a variable downsampling frequency for the embedding of the trajectories $X_{i-}$ and $X_{j-}$, and an adjustable correlation exclusion window to obtain less correlated samples of the data $(X_{j+},X_{i-},X_{j-})$, aside from the primary hyperparameter choices of the number of nearest neighbors $k^{'}$ and the choice of a distance metric to compute nearest neighbors in the trajectory space\cite{smvicente2011transfer}. It has, in fact, been argued that the absolute value of the transfer entropy obtained via the KSG estimator in practice depends so heavily on the choices of these hyperparameters that it may not be any more reliable than the value merely being zero or non-zero, \textit{i.e.}, for testing for statistical independence\cite{smwibral2014directed}.

In order to compare the performance of the KSG algorithm with that of TE-PWS on a fair footing, we have performed an extensive search through the hyperparameter space of the KSG algorithm to obtain converged transfer entropy rate estimates. For all KSG computations, we keep the number of nearest neighbors fixed at $k^{'}=4$, which was previously reported to be optimal\cite{smkraskov2004estimating}, and the choice of distance metric to be the $L_{\infty}$ norm, which is already implemented in the JIDT toolkit\cite{smlizier2014jidt}. Additionally, since the KSG estimate is only known to be unbiased for independent data samples, while samples obtained from a trajectory are inherently temporally correlated, we choose a dynamical correlation exclusion window of $\tau$ equal to the relaxation timescale of the system, a choice previously reported to be optimal\cite{smvicente2011transfer,smlizier2014jidt}. Fig. \ref{fig:ksg} then shows our protocol for the nonlinear model D. First, we find an optimal \textit{downsampling} rate such that the transfer entropy rate is not unphysically negative. For this we fix the total duration of trajectory data, $NM_{1}\delta t$, to be $1.6\times10^{6}\delta t_{0}$, where $\delta t_{0}$ is the simulation timestep. We also fix, at this stage, the history length to a value slightly larger than 1, $k+1=5$. We then downsample the simulated fine-grained trajectory with a timestep $\delta t$, a multiple of the simulation timestep $\delta t_{0}$. This means that we keep every $(\delta t/\delta t_{0})$-th frame and discard the other frames from the trajectories. We then compute the transfer entropy rate $\dot{\mathcal{T}}_{X_{1}\to X_{3}}$ with the downsampled trajectories with $k+1=5$ for each value of $\delta t$. The result is plotted in Fig. \ref{fig:ksg}a. We find that if we do not downsample at all, \textit{i.e.}, $\delta t=\delta t_{0}$, the transfer entropy rate estimate from the KSG algorithm becomes negative, which is unphysical and is a previously reported numerical artifact of the algorithm\cite{smholmes2019estimation,smmarx2022estimating}. The artifact in this limit is likely to be the result of a strong correlation between the consecutive frames in the simulated trajectories, since the KSG algorithm is known to have a large systematic error in the presence of strongly correlated data\cite{smgao2015efficient}. On the other hand, a converged KSG estimate has been reported for experimental data.\cite{smlizier2014jidt} This observation could be rationalized by noting that the experimental time series data is inherently sampled at an interval that is larger than the timestep used in the simulations. This does imply, however, that the successive time points in the experimental data appear less correlated, meaning that a part of the transfer entropy will be missed in the KSG estimate. We see this effect in Fig. \ref{fig:ksg}a where the transfer entropy estimate gradually decays to zero at higher downsampling rates $\delta t$. As a compromise between the two uncontrolled errors, we choose multiple values of $\delta t$ that are close to the peak in Fig. \ref{fig:ksg}a for optimizing the other hyperparameters.

\begin{figure}[t]
\centering
\includegraphics[width=18cm]{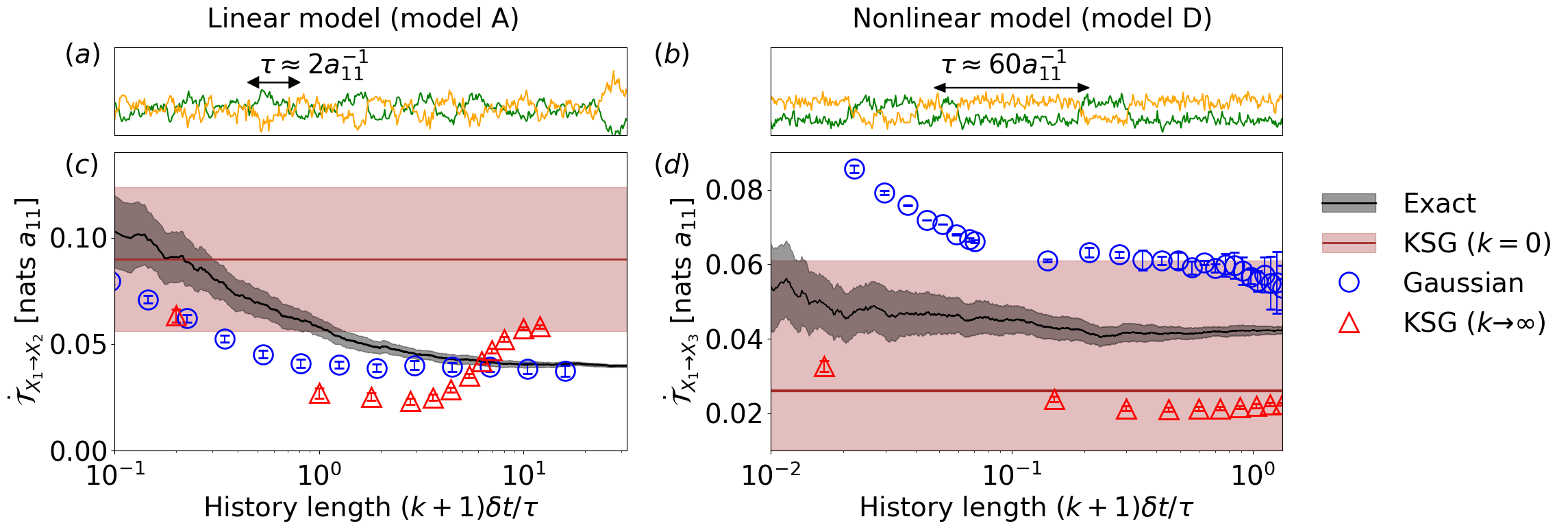}
\caption{Comparison of $\dot{\mathcal{T}}$ estimates from different widely used approximate methods with the exact results from TE-PWS in a linear and a nonlinear model. (a) Trajectories of the source and sink variables $X_{1}$ (green) and $X_{2}$ (orange), respectively, in the linear model A with $\tau$ denoting the relaxation timescale. (b) Trajectories of the source and sink variables $X_{1}$ (green) and $X_{3}$ (orange), respectively, in the nonlinear model D with $\tau$ as the relaxation timescale. (c) Convergence of the estimates of $\dot{\mathcal{T}}_{X_{1}\to X_{2}}$ from different methods as a function of the history length divided by the relaxation timescale of the dynamics. The Gaussian and KSG $(k\to\infty)$ points were obtained through an extensive scanning of hyperparameters for physically meaningful converged values in the infinite data limit, see Figs. \ref{fig:gaussian} and \ref{fig:ksg} for the protocol. The exact estimate was obtained from TE-PWS using Eq. 3 of the main text by summing over all $k$ and dividing by the total trajectory duration. The KSG $(k=0)$ estimate was obtained with the KSG algorithm with 4 nearest neighbors by extrapolating to the infinite data limit ($1/(NM_{1})\to 0$) from three different values of $NM_{1}$, which are $NM_{1}=1.6\times 10^{6}$ and half and one-third of that data. Unlike Fig. 2(a) of the main text where we plot $\dot{\mathcal{T}}^{[k]}_{X_{1}\to X_{2}}$ for \textit{both} methods, here we plot $\dot{\mathcal{T}}^{[k]}_{X_{1}\to X_{2}}$ for the Gaussian framework and $\dot{\mathcal{T}}_{X_{1}\to X_{2}}=\left(\sum_{k=0}^{N-1}\mathcal{T}^{[k]}_{X_{1}\to X_{2}}\right) /(N\delta t)$ for TE-PWS (see Eq. 3 of the main text). The Gaussian framework and the TE-PWS estimates thus differ for intermediate $k$, but converge in the large $k$-limit, yielding the desired transfer entropy rate. In contrast, the KSG method gives an inaccurate estimate in the large $k$ limit. (d) Similar to (c) for the estimate of $\dot{\mathcal{T}}_{X_{1}\to X_{3}}$ in the nonlinear model. For obtaining the KSG $(k\to\infty)$ data points, downsampling rates of 40 and 100 have been used in the linear and the nonlinear models, respectively. Error bars in each estimate are evaluated from 40 independent realizations.}
\label{fig:compare}
\end{figure}

Panels (b), (c) and (d) in Fig. \ref{fig:ksg} show the subsequent hyperparameter optimization for three choices of the downsampling time step $\delta t$: $\delta/\delta t_{0}= 1$, $50$ and $100$. For each given $\delta t$, we first choose a fixed history length $(k+1)$, compute the transfer entropy for different data sizes $NM_{1}=1.28\times 10^{5}$, $1.28\times 10^{5}/2$ and $1.28\times 10^{5}/3$, and linearly extrapolate to the infinite data limit $1/(NM_{1})\to0$. We then perform this extrapolation at increasing values of $k$ and try to obtain an extrapolated result converged with respect to $k$, \textit{i.e.}, independent of $k$. Fig. \ref{fig:ksg}b shows that, without downsampling, even this careful extrapolation cannot correct the unphysically negative values of the KSG estimate, as the extrapolated values (black symbols in Fig. \ref{fig:ksg}b) continue to remain negative for increasing $k$. This implies that, without downsampling, the required data size by the algorithm to give a physically meaningful answer in this problem is so large that our calculated estimates are not even in the asymptotic $\sim 1/(NM_{1})$ scaling limit. Fig. \ref{fig:ksg}c shows that this problem persists even if we choose a downsampling rate at the peak of Fig. \ref{fig:ksg}a, \textit{i.e.}, $\delta t/\delta t_{0}=50$. Even though the transfer entropy rate estimates here are non-negative, they do not converge in the large history length limit after linearly correcting for the finite data size. Only after choosing an even larger downsampling rate, $\delta t/\delta t_{0}=100$, we are able to obtain a result converged with respect to both data size and history length as Fig. \ref{fig:ksg}d shows. This converged estimate and the associated computational cost has been described in Table 1 of the main text. The protocol is derived similarly for the linear system (model A), where the downsampling rate has been chosen to be $\delta t/\delta t_{0}=40$, $NM_{1}$ values have been varied over $1.6\times 10^{5}$, $1.6\times 10^{5}/2$ and $1.6\times 10^{5}/3$ to extrapolate to the $1/(NM_{1})\to 0$ limit, and $k$ has been varied for convergence up to $k\delta t/\delta t_{0}=2400$. For the sake of completeness, we mention here that we have also varied other hyperparameters such as a downsampling (embedding) frequency for only the trajectories $X_{i-}$ and $X_{j-}$ (data not shown) following the protocol described in Ref. \citen{smvicente2011transfer}, in order to substitute the downsampling of the entire trajectories of $X_{i}$ and $X_{j}$. However, we faced the same problem of an unphysical negative value of the transfer entropy rate estimate in those cases as well.

\textbf{Summary.} Fig. \ref{fig:compare} summarizes the convergence of the transfer entropy rate estimates from all methods in the linear and nonlinear models. The TE-PWS estimates use the $NM_{1}$ values mentioned in Table 1 of the main text, and $M_{2}=100$ in all cases. The Gaussian and KSG ($k\to\infty$) estimates were obtained as described in the above paragraphs and are the same results as mentioned in Table 1 of the main text. Additionally, we have also computed the one-step transfer entropy using the KSG algorithm, labeled as KSG ($k=0$). For this estimate, downsampling has not been necessary. We obtained an estimate independent of data-size by using $NM_{1}=1.6\times 10^{6}$, $1.6\times 10^{6}/2$ and $1.6\times 10^{6}/3$, and linearly extrapolating to the infinite data limit. Fig. \ref{fig:compare} demonstrates that all approximate techniques incur significant systematic errors in at least one case. In particular, while the Gaussian framework is accurate for the linear system, as expected, it fails to accurately predict the transfer entropy for the nonlinear system. The KSG ($k\to\infty$) result significantly deviates from the exact TE-PWS result for both the linear and nonlinear systems. The one-step transfer entropy estimated using KSG, KSG ($k=0$), is inaccurate for the linear system and imprecise for the nonlinear system. In summary, the advantages of the exact estimate from TE-PWS compared to widely used approximate methods are as follows:
\begin{itemize}
\item Depending on the amount of nonlinearity and feedback in the dynamics, all approximate methods can incur large systematic errors in the transfer entropy rate estimate.
\item All approximate methods need to be corrected for large sample-size dependent biases at larger values of the history length $(k+1)$. In contrast, the TE-PWS estimate is unbiased at any given sample-size. The only effect of a smaller sample-size on the TE-PWS estimate is larger statistical error bars.
\item In order to obtain the convergence of $\dot{\mathcal{T}}_{X_{i}\to X_{j}}^{[k]}$ with respect to increasing $k$, all approximate methods require a new computation for every new value of $k$. In contrast, TE-PWS gives the $k$-dependence of $\dot{\mathcal{T}}_{X_{i}\to X_{j}}^{[k]}$ for all values of $k$ in one simulation.
\item As shown in Table 1 of the main text, TE-PWS has either comparable or orders of magnitude lower computational cost than the approximate methods.
\end{itemize}



\section{SM-I: Data processing inequality for information}
The Data Processing Inequality (DPI) constrains the mutual information rate between the trajectories of an information source and sink in the presence of a mediator variable\cite{smcover1999elements}. Specifically, the DPI for the motif $X_{1}\rightleftarrows X_{2}\to X_{3}$, studied in the main text, can be derived by starting from the relation $I(X_{1,[0,N]};X_{3,[0,N]}|X_{2,[0,N]})=0$, \textit{i.e.}, the statistical independence of the source and sink trajectories when conditioned on the trajectory of the mediator variable. Denoting henceforth $I(X_{i,[0,N]};X_{j,[0,N]})$ as $I(X_{i};X_{j})$, which, to emphasize, refers to the mutual information between the entire trajectories of $X_{i}$ and $X_{j}$ and not the instantaneous values, we then use the chain rule of mutual information two times,
\begin{align}
I(X_{1};X_{3}|X_{2})=0=I(X_{1};X_{2},X_{3})-I(X_{1};X_{2})=I(X_{1};X_{2}|X_{3})+I(X_{1};X_{3})-I(X_{1};X_{2})
\end{align}
which implies, by the non-negativity of the mutual information $I(X_{1};X_{2}|X_{3})$,
\begin{align}
I(X_{1};X_{3})=I(X_{1};X_{2})-I(X_{1};X_{2}|X_{3})\leq I(X_{1};X_{2})
\end{align}
which is the DPI. The corresponding mutual information rates, defined as $\dot{I}(X_{i};X_{j})\equiv\lim_{N\to\infty} I(X_{i};X_{j})/(N\delta t)$, thus also obey the DPI $\dot{I}(X_{1};X_{3})\leq \dot{I}(X_{1};X_{2})$.

\begin{figure}[H]
\centering
\includegraphics[width=18cm]{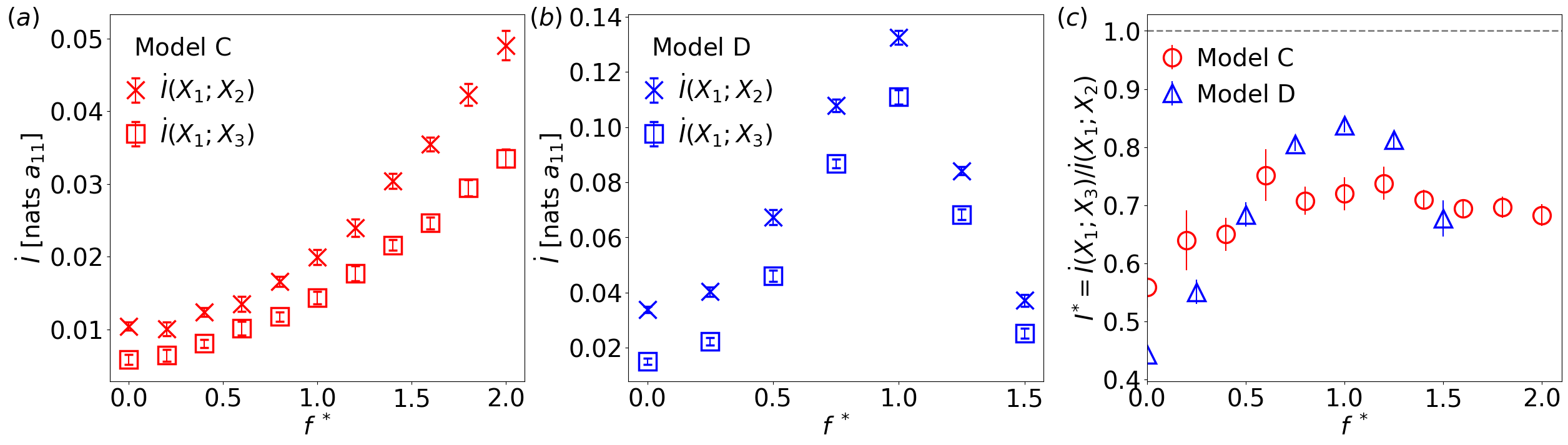}
\caption{(a) and (b) show mutual information rates as a function of increasing feedback $f^{*}$ from $X_{2}$ to $X_{1}$ in the linear model C and the nonlinear model D respectively, for the motif $X_{1}\rightleftarrows X_{2}\to X_{3}$. (a) In the linear model C, both the mutual information rates $\dot{I}(X_{1};X_{2})$ and $\dot{I}(X_{1};X_{3})$ monotonically increase with $f^{*}$. This is in contrast to the results in Fig. 3c of the main text, where we showed that the transfer entropy rate $\dot{\mathcal{T}}^{(\mathrm{C})}_{X_{1}\to X_{2}}$ stays constant with increasing $f^{*}$. (b) In the nonlinear model D, both mutual information rates peak around the switching regime $f^{*}=1$. (c) The ratio of the mutual information rates, $I^{*}=\dot{I}(X_{1;X_{3}})/\dot{I}(X_{1};X_{2})$, stays below the Data Processing Inequality bound of unity (black dashed line) for both models.}
\label{fig:infodpi}
\end{figure}

We demonstrate this numerically in Fig. \ref{fig:infodpi} with exact results from the PWS algorithm\cite{smreinhardt2023path}. Panels (a) and (b) of Fig. \ref{fig:infodpi} show the mutual information rates $\dot{I}(X_{1};X_{2})$ and $\dot{I}(X_{1};X_{3})$ for models C and D, respectively, as a function of $f^{*}$, the strength of the $X_{2}\rightarrow X_{1}$ feedback relative to the $X_{1}\rightarrow X_{2}$ coupling. In model C, both quantities monotonically increase with increasing $f^{*}$ because $X_{2}$ and subsequently $X_{3}$ become more strongly correlated with $X_{1}$. In model D, both information rates have a peak at the switching regime, around $f^{*}=1$. 
However, unlike the transfer entropy rates shown in Fig. 3 of the main text, the mutual information rates here continue to obey the DPI, as shown in Fig. \ref{fig:infodpi}c, with the ratio $I^{*}=\dot{I}(X_{1};X_{3})/\dot{I}(X_{1};X_{2})$ being always smaller than unity. Importantly, each mutual information rate is bounded from below by the sum of the corresponding forward and backward transfer entropy rates\cite{smchetrite2019information}, for example, $\dot{I}(X_{1};X_{3})\geq\dot{\mathcal{T}}_{X_{1}\to X_{3}}+\dot{\mathcal{T}}_{X_{3}\to X_{1}}$ and $\dot{I}(X_{1};X_{2})\geq\dot{\mathcal{T}}_{X_{1}\to X_{2}}+\dot{\mathcal{T}}_{X_{2}\to X_{1}}$. When the $X_{2}\rightarrow X_{1}$ feedback strength $f^{*}$ is increased, even though $\dot{\mathcal{T}}_{X_{1}\to X_{3}}$ increases faster than $\dot{\mathcal{T}}_{X_{1}\to X_{2}}$, the backward transfer entropy $\dot{\mathcal{T}}_{X_{2}\to X_{1}}$ that directly quantifies the feedback increases even faster compared to $\dot{\mathcal{T}}_{X_{3}\to X_{1}}$. As a result, $\dot{I}(X_{1};X_{2})$ stays above $\dot{I}(X_{1};X_{3})$, restoring the DPI for information.

\end{document}